\documentclass[secnumarabic, nobibnotes, aps, prd]{revtex4-2}
\usepackage{amssymb,epsf}
\usepackage{graphicx}
\usepackage{amsmath}
\usepackage{subcaption}
\usepackage{multirow}
\usepackage{adjustbox}

\begin{document}

\title{Thermodynamic topology of topological charged dilatonic black holes}
\author{B. Hazarika$^{1}$\footnote{%
email address: bidyuthazarika1729@gmail.com}, B. Eslam Panah$^{2}$ \footnote{%
email address: eslampanah@umz.ac.ir}, and P. Phukon$^{1,3}$\footnote{%
email address: prabwal@dibru.ac.in}}

\affiliation{$^{1}$ Department of Physics, Dibrugarh University, 786004, Dibrugarh, Assam, India\\
$^{2}$ Department of Theoretical Physics, Faculty of Basic Sciences, University of Mazandaran, P. O. Box 47416-95447, Babolsar, Iran\\
$^{3}$ Theoretical Physics Division, Centre for Atmospheric Studies, Dibrugarh University, Dibrugarh, Assam, India}

\begin{abstract}
The aim of this paper is to explore the thermodynamic topology of
topological charged dilatonic black holes. To achieve this, our study will
begin by examining the characteristics of topological charged black holes in
dilaton gravity. Specifically, we will concentrate on the impact of the
topological constant on the event horizon of these black holes.
Subsequently, we will analyze these black holes, considering their
thermodynamic and conserved quantities, in order to assess the validity of
the first law of thermodynamics. We explore the thermodynamic topology of
these black holes by treating them as thermodynamic defects. For our study,
we examine two types of thermodynamic ensembles: the fixed $q$ ensemble and
the fixed $\phi$ ensemble. To study the impact of the topological constant ($%
k$) on thermodynamic topology, we consider all possible types of curvature
hypersurfaces that can form in these black holes. By calculating the
topological charges at the defects within their thermodynamic spaces, we
analyze both the local and global topology of these black holes. We also
investigate how the parameters of dilaton gravity affect the thermodynamic
topology of black holes and highlight the differences compared to charged
black holes in the General Relativity.
\end{abstract}

\maketitle

\section{Introduction}

Despite many successes in agreement with observations, General Relativity
(GR) cannot explain some phenomena such as the cosmological constant problem 
\cite{Weinberg1989}, the origin of acceleration of our universe \cite%
{Riess1998,Perlmutter1999}, cosmic microwave background (CMB) radiation \cite%
{Spergel2003,Ade2016}, and the existence of dark energy and dark matter. So
we need to modify this theory of gravity. On the other hand, dark energy and
especially dark matter have received much attention recently. In theoretical
physics, non-baryonic dark matter has been classified into three models:
cold, warm, and hot \cite{Dick1996}. The dilaton field is one of the most
fascinating candidates for cold dark matter because the cold dark matter
model agrees well with experimental observations \cite{Cho1990}. In
addition, considering a new scalar field is one of the best ways to find the
nature of dark energy \cite{Huang2007,Huang2008}. It is notable that, the
low energy limit of string theory contains a dilaton field that is coupled
to gravity. The dilaton coupling with other gauge fields profoundly affects
the resulting solutions \cite%
{Koikawa1987,Gibbons1988,Brill1991,Garfinkle1991,Horowitz1991,Gregory1993}.
For example, it was indicated that a dilaton field can change the asymptotic
behavior of the spacetime. In particular, in the presence of one or two
Liouville-type dilaton potentials, black hole spacetimes are neither
asymptotically flat nor (anti)-de Sitter ((A)dS) \cite%
{Mignemi1992,Poletti1995,CaiJS1998,Clement2003,Hendi2008}. This is because
the dilaton field does not vanish for $r\rightarrow \infty $. Furthermore,
it was shown that the combination of three Liouville-type dilaton potentials
allows the construction of dilaton black hole solutions on the (A)dS
spacetime background \cite{Gao2005,Hendi2010}. It was also shown that the
dilaton charges are expressed as black hole masses and the scalar dilaton
field acts as a secondary hair. Furthermore, the dilaton field could change
the causal structure of the black hole and lead to the curvature
singularities at finite radii \cite%
{Mignemi1992,Poletti1995,CaiJS1998,Clement2003}. Therefore, adding a dilaton
field to GR to investigate the physical properties of black holes has been
getting more attention \cite%
{TamakiT2000,YamazakiI2001,Yazadjiev2005,Dehghani2008,LiuLN2015,HendiFEP2016,MoLX2016,HendiEP2016,QuevedoQS2016,LiSZ2018,EslamHPH2018,AzregHJR2019,BritoP2018,Stetsko2019,SalcedoKK2019,DehyadegariS2020,Junior2021,Zhang2023,Boshkayev2024}%
.

From a quantum perspective, black holes are capable of emitting energy, a
phenomenon known as Hawking radiation. This leads us to consider black holes
as thermal objects with a specific temperature \cite{Hawking1974,Hawking1975}%
. Furthermore, black holes possess a measurable entropy, referred to as the
Hawking-Bekenstein entropy, which can be derived from their surface area 
\cite{Bekenstein1973,Bekenstein1974}. Building upon these concepts, Bardeen
et al. introduced the thermodynamics of black holes \cite{BardeenCH1973}. In
this regard, researchers have dedicated numerous years to studying the
thermodynamics of various types of black holes. Some investigations have
focused on the extended phase space using the anti-de Sitter/Conformal Field
Theory (AdS/CFT) correspondence \cite{Witten1998,RyuT2006}. In this
approach, the cosmological constant is treated as pressure \cite{Kastor2009}%
, offering valuable insights into black hole phase transitions \cite%
{Kubiznak2012,Altamirano2013,Ther1,Ther2,Ther3,Ther4,Ther5,Ther6,Ther7,Ther8,Ther9}%
.

The study of black hole thermodynamics has generated significant interest in
the field of topology. Two topological approaches have been developed to
investigate the thermodynamic properties of various black hole systems. A
topological concept was introduced in Ref. \cite{WeiL2022}, which involves a
vector with zero points that correspond to the critical point. According to
Duan's pioneering $\phi$-mapping topological current theory \cite%
{Duan1979,Duan1984}, these zero points represent defects of the vector, as
the direction of the vector is undefined at these points. From a topological
standpoint, each defect can be assigned a winding number, which is an
integer value. By summing all the winding numbers, a topological number is
obtained, allowing for the classification of the systems into different
topological classes. Within each class, the systems share similar
thermodynamic properties. Specifically, based on the value of the winding
number, the critical point is divided into two classes: the conventional
class and the novel class. These classes exhibit opposite properties in
terms of the generation and annihilation of the black hole branches.
Furthermore, the black hole systems are grouped into different classes based
on the structures of the critical points. In this regard, the thermodynamic
topology of black holes has been studied by considering various black holes
in much literature \cite%
{TheTo1,TheTo2,TheTo3,TheTo4,TheTo5,TheTo6,TheTo7,TheTo8,TheTo9,TheTo10,TheTo11,TheTo12,TheTo13,TheTo14,TheTo15,TheTo16,TheTo17,TheTo18,TheTo19}%
. In this paper, we follow the topological approach proposed in Ref. \cite%
{TheTo2}, and study the thermodynamic topology by treating the charged
dilatonic black holes as topological defects.

This paper is organized as follows. In Section II, we derive the solution
for topological charged black holes in dilaton gravity. Section III
discusses the thermodynamic properties and conserved quantities associated
with this solution. Section IV explores the concept of thermodynamic
topology and outlines the mathematical procedures needed for its
investigation. Subsections IV.1 and IV.2 focus on two statistical ensembles:
the fixed charge ($q$) ensemble and the fixed potential ($\phi$) ensemble.
Within these subsections, we examine various types of curvature
hypersurfaces to understand how the topological constant ($k$) impacts
thermodynamic topology. Finally, section V provides our conclusions and
closing remarks on the study.

\section{Black hole solution}

The action of charged dilaton gravity is \cite{Chan} 
\begin{equation}
\mathcal{I}=\frac{1}{16\pi }\int d^{4}x\sqrt{-g}\left[ \mathcal{R}-2\left(
\nabla \Phi \right) ^{2}-V\left( \Phi \right) -e^{-2\alpha \Phi }F_{\mu \nu
}F^{\mu \nu }\right] ,  \label{action}
\end{equation}%
in the above action, $\mathcal{R}$ represents the Ricci scalar curvature, $%
\Phi $ represents the dilaton field, and $V\left( \Phi \right) $ represents
the potential for $\Phi $. The electromagnetic field is denoted by $F_{\mu
\nu }=\partial _{\mu }A_{\nu }-\partial _{\nu }A_{\mu }$, with $A_{\mu }$
representing the electromagnetic potential. It is important to note that $%
\alpha $ is a constant that determines the strength of the coupling between
the scalar and electromagnetic fields. For our purposes of finding a black
hole with a radial electric field ($F_{tr}(r)=-F_{rt}(r)\neq 0$), the form
of the electromagnetic potential will be as follows 
\begin{equation}
A_{\mu }=\delta _{\mu }^{0}h\left( r\right) .  \label{electric po}
\end{equation}

By utilizing the variational principle and by varying Eq. (\ref{action})
with respect to the gravitational field $g_{\mu \nu}$, the dilaton field $%
\Phi$, and the gauge field $A_{\mu}$, we are able to obtain the subsequent
field equations \cite{Chan}. 
\begin{eqnarray}
R_{\mu \nu } &=&2\left( \partial _{\mu }\Phi \partial _{\nu }\Phi +\frac{1}{4%
}g_{\mu \nu }V(\Phi )\right) +2e^{-2\alpha \Phi }\left( F_{\mu \eta }F_{\nu
}^{\eta }-\frac{1}{4}g_{\mu \nu }F_{\lambda \eta }F^{\lambda \eta }\right) ,
\label{dilaton equation(I)} \\
&&  \notag \\
\nabla ^{2}\Phi &=&\frac{1}{4}\frac{\partial V}{\partial \Phi }-\frac{\alpha 
}{2}e^{-2\alpha \Phi }F_{\lambda \eta }F^{\lambda \eta },
\label{dilaton equation(II)} \\
&&  \notag \\
\nabla _{\mu }\left( e^{-2\alpha \Phi }F^{\mu \nu }\right) &=&0.
\label{Maxwell equation}
\end{eqnarray}

The objective of this paper is to obtain topological charged black hole
solutions from Einstein-Maxwell-dilaton gravity. To accomplish this, we will
employ the following static metric ansatz 
\begin{equation}
ds^{2}=-f(r)dt^{2}+\frac{dr^{2}}{f(r)}+r^{2}R^{2}(r)d\Omega _{k}^{2},
\label{metric}
\end{equation}%
where the functions $f(r)$ and $R(r)$ must be determined, while $d\Omega
_{k}^{2}$ represents the line element of a $2-$dimensional hypersurface with
a constant curvature of $2k$ and volume $\mathcal{V}$. It is important to
note that the constant $k$ indicates that the boundary of $t=constant$ and $%
r=constant$ can be a positive (elliptic), zero (flat), or negative
(hyperbolic) constant curvature hypersurface, expressed explicitly as
follows 
\begin{equation}
d\Omega _{k}^{2}=\left\{ 
\begin{array}{cc}
d\theta ^{2}+\sin ^{2}\theta d\varphi ^{2}, & k=+1 \\ 
d\theta ^{2}+\sinh ^{2}\theta d\varphi ^{2}, & k=-1 \\ 
d\theta ^{2}+d\varphi ^{2}, & k=0%
\end{array}%
\right. .  \label{dOmega}
\end{equation}

by utilizing Eqs. (\ref{electric po}), (\ref{Maxwell equation}) and (\ref%
{metric}), it is possible to derive the electromagnetic tensor 
\begin{equation}
F_{tr}=\frac{qe^{2\alpha \Phi \left( r\right) }}{r^{2}R^{2}(r)},
\label{Ftr eq}
\end{equation}%
where $q$ is an integration constant that is associated with the electric
charge of the black hole.

Here, we employ an adjusted form of a Liouville-type dilation potential to
discover metric functions that are consistent. The modified version of the
potential has the following form \cite{HendiFEP2016} 
\begin{equation}
V(\Phi )=\frac{2k\alpha ^{2}}{b^{2}\mathcal{K}_{1,-1}}e^{\frac{2\Phi \left(
r\right) }{\alpha }}+2\Lambda e^{2\alpha \Phi \left( r\right) },
\label{V(Phi)}
\end{equation}%
where $\Lambda $ is a free parameter which plays the role of the
cosmological constant. Also, we define $\mathcal{K}_{i,j}=i\alpha ^{2}+j$,
for example, $\mathcal{K}_{1,-1}=\alpha ^{2}-1$. It is notable that, the
Liouville-type dilation potential, denoted as $V(\Phi )$, is utilized in the
study of Einstein-Maxwell-dilaton black holes \cite{Yazadjiev2005}, and
Friedman-Robertson-Walker scalar field cosmologies \cite{Ozer1992}.

Next, by applying the ansatz $R(r)=e^{\alpha \Phi (r)}$ and substituting Eq.
(\ref{Ftr eq}) into the field equations (Eqs. (\ref{dilaton equation(I)})
and (\ref{dilaton equation(II)})), we can obtain the following solutions in
Einstein-Maxwell-dilaton gravity \cite{Sheykhi2007,HendiSPE2015} 
\begin{eqnarray}
f(r) &=&-\frac{\mathcal{K}_{1,1}}{\mathcal{K}_{1,-1}}k\left( \frac{b}{r}%
\right) ^{\frac{-2\alpha ^{2}}{\mathcal{K}_{1,1}}}-\frac{m}{r^{\frac{%
\mathcal{K}_{-1,1}}{\mathcal{K}_{1,1}}}}+\frac{\mathcal{K}_{1,-1}^{2}\Lambda
r^{2}}{\mathcal{K}_{1,-3}}\left( \frac{b}{r}\right) ^{\frac{2\alpha ^{2}}{%
\mathcal{K}_{1,1}}}+\frac{q^{2}\mathcal{K}_{1,1}}{r^{2}}\left( \frac{b}{r}%
\right) ^{\frac{-2\alpha ^{2}}{\mathcal{K}_{1,1}}},  \label{f(r)} \\
&&  \notag \\
\Phi (r) &=&\frac{\alpha }{\mathcal{K}_{1,1}}\ln \left( \frac{b}{r}\right) ,
\label{Phi(r)}
\end{eqnarray}%
where $b$ is an arbitrary constant. Additionally, in the above expression, $%
m $ is an integration constant that is associated with the geometrical mass
of the black hole. It is worth noting that when the parameters of dilaton
gravity, i.e. $\alpha =0$, are absent, the solution (\ref{f(r)}) becomes a $%
4 $-dimensional asymptotically AdS topological charged black hole in GR with
a positive, zero, or negative constant curvature hypersurface, as $f(r)=k-%
\frac{m}{r}-\frac{\Lambda }{3}r^{2}+\frac{q^{2}}{r^{2}}$.

We are searching for the curvature singularity to validate the
interpretation of the solutions as black holes. Therefore, we calculate the
Kretschmann scalar. The calculations indicate that the Kretschmann scalar is
finite for finite values of the radial coordinates. However, for extremely
small and large values of $r$, we obtain the following results 
\begin{eqnarray}
\lim_{r\rightarrow 0}R_{\alpha \beta \mu \nu }R^{\alpha \beta \mu \nu }
&\propto &r^{-\frac{4\mathcal{K}_{1,2}}{\mathcal{K}_{1,1}}},  \label{RR0} \\
&&  \notag \\
\lim_{r\rightarrow \infty }R_{\alpha \beta \mu \nu }R^{\alpha \beta \mu \nu
} &=&\frac{12(\alpha ^{4}-2\alpha ^{2}+2)\Lambda }{\mathcal{K}_{-1,3}^{2}}%
\left( \frac{b}{r}\right) ^{\frac{4\alpha ^{2}}{\mathcal{K}_{1,1}}},
\label{RRinf}
\end{eqnarray}%
Equation (\ref{RR0}) confirms the presence of an essential singularity at $%
r=0$, while Eq. (\ref{RRinf}) demonstrates that, for nonzero $\alpha $, the
solutions do not exhibit asymptotically flat or (A)dS behavior.

\begin{figure}[tbh]
\centering
\includegraphics[width=0.4\textwidth]{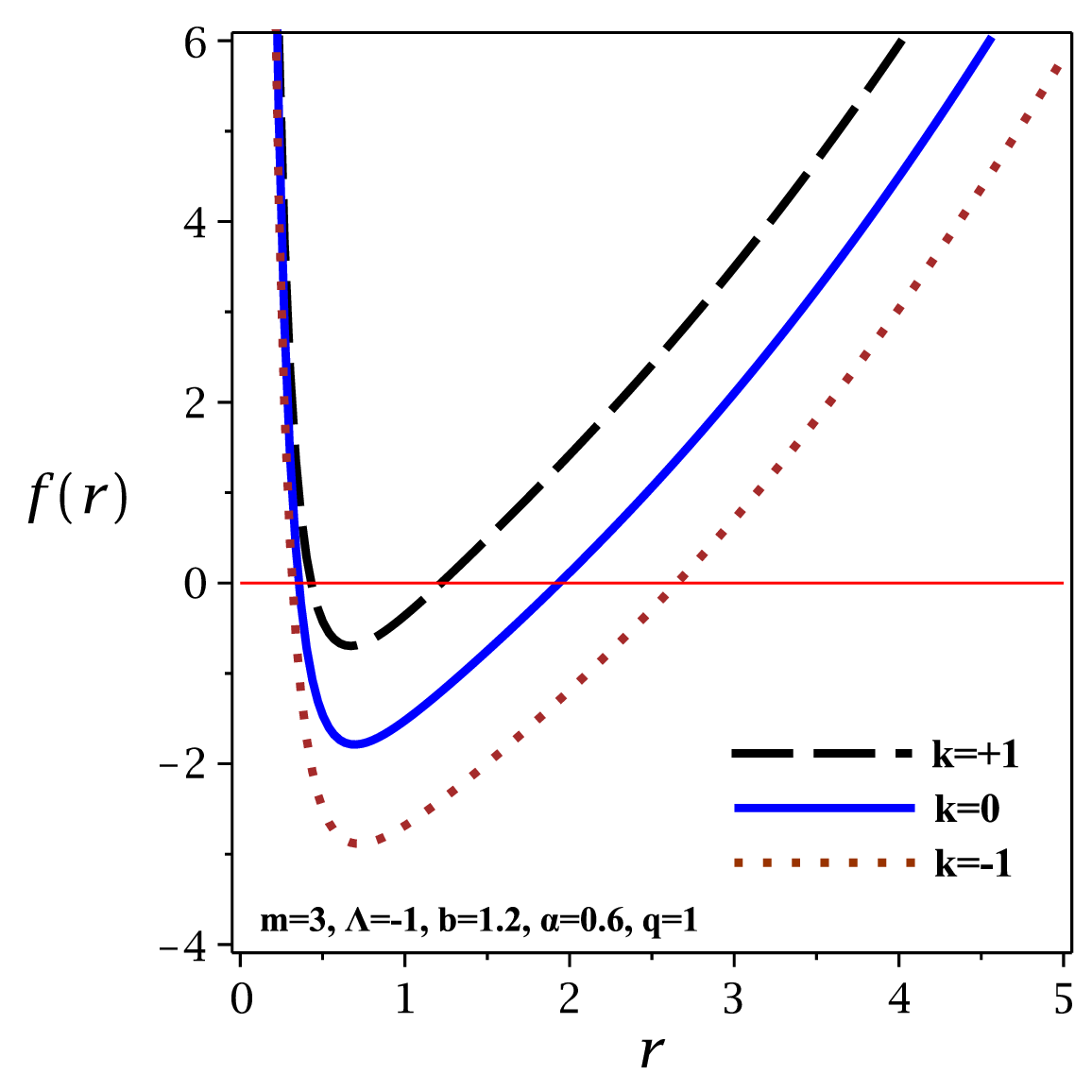} \newline
\caption{$f(r)$ versus $r$ for different values of $k$.}
\label{Fig1}
\end{figure}

\begin{figure}[tbh]
\centering
\includegraphics[width=0.35\textwidth]{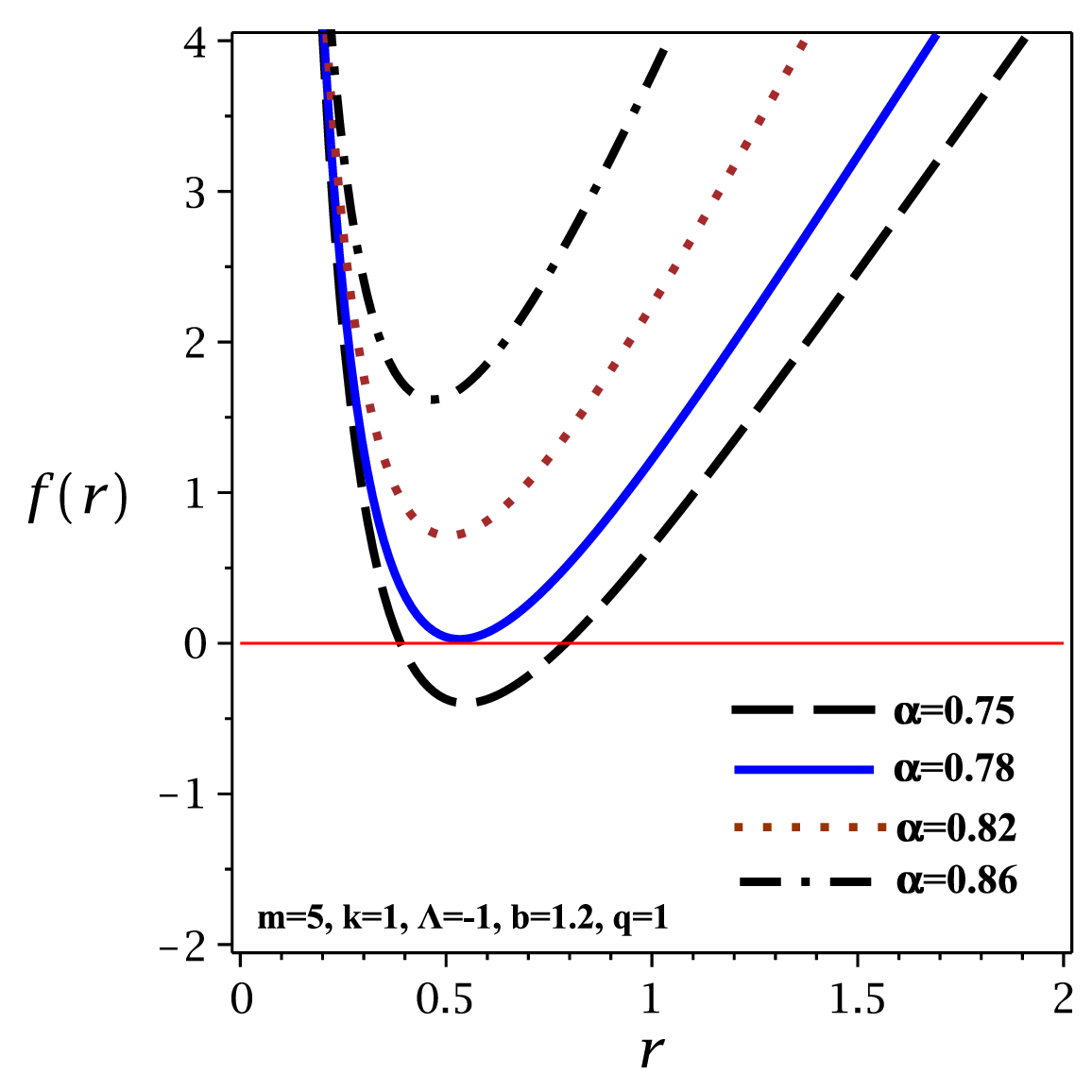} %
\includegraphics[width=0.35\textwidth]{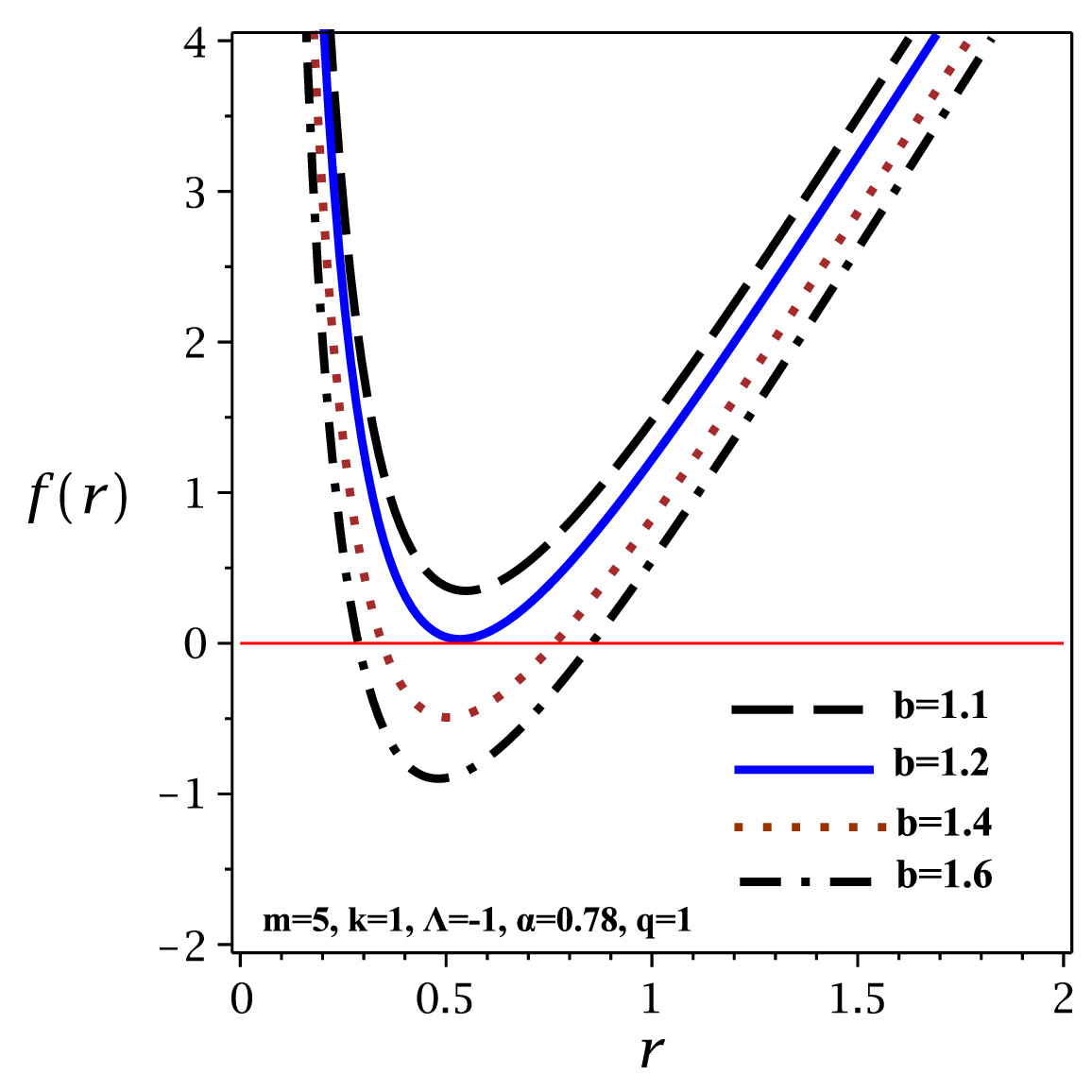}\newline
\includegraphics[width=0.35\textwidth]{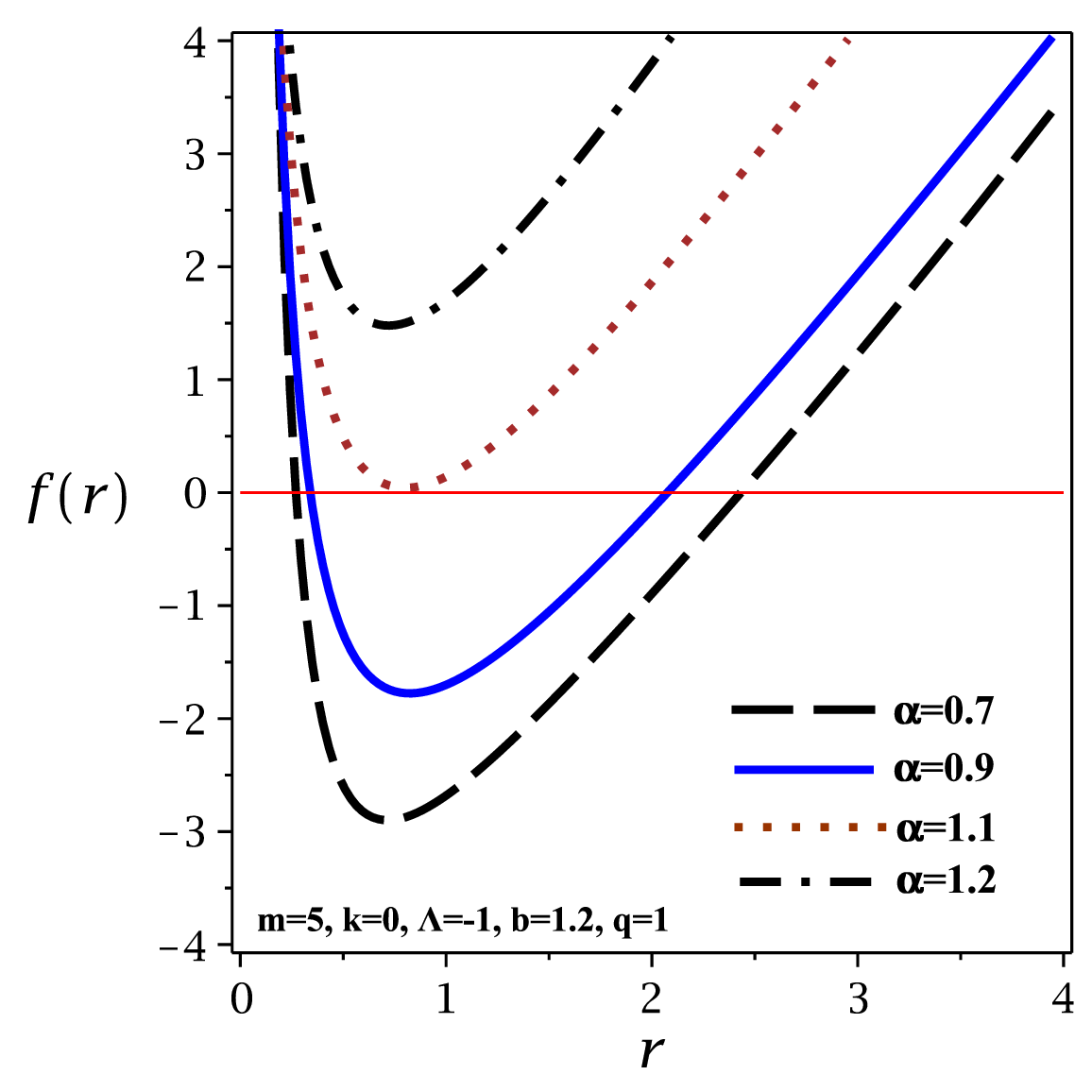} %
\includegraphics[width=0.35\textwidth]{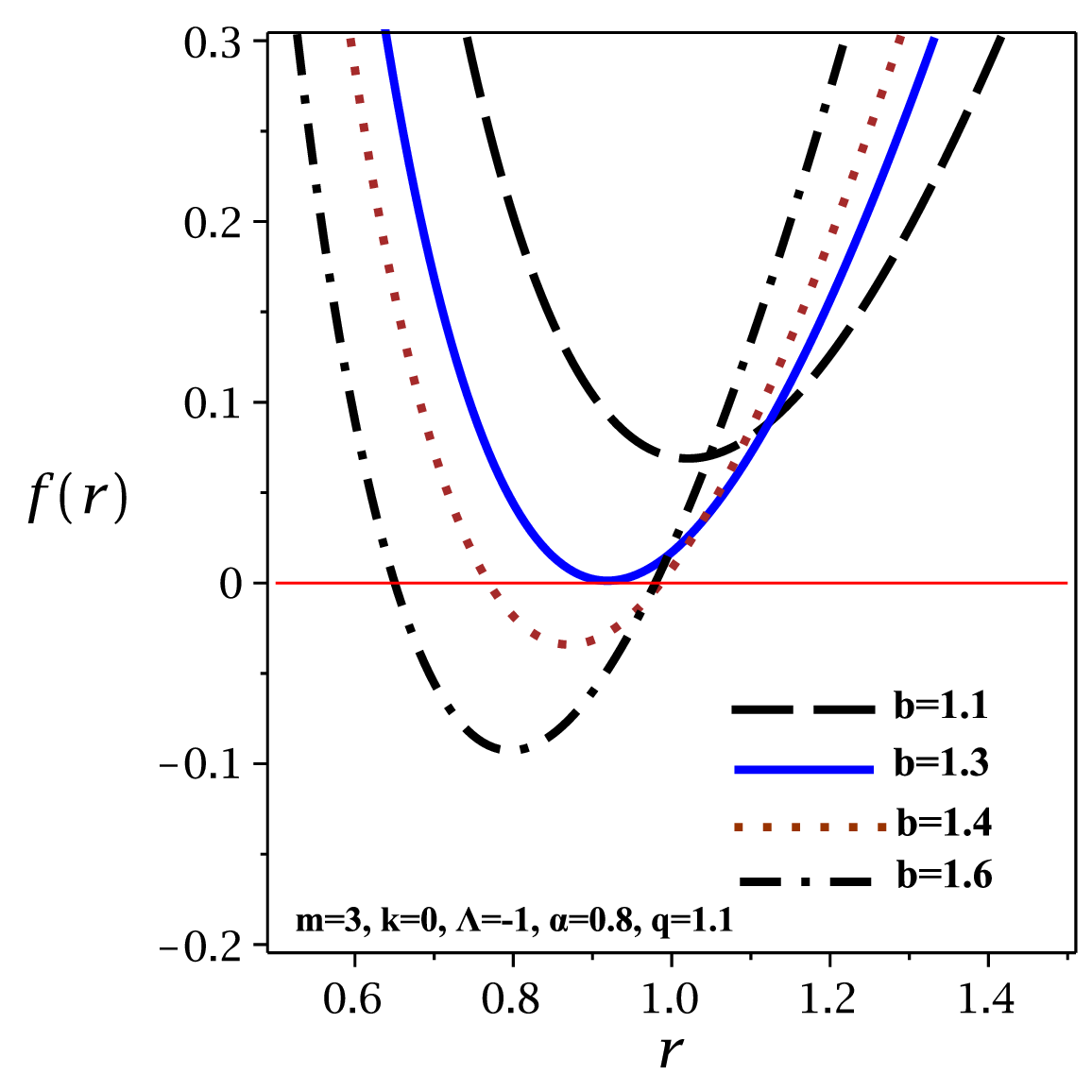}\newline
\includegraphics[width=0.35\textwidth]{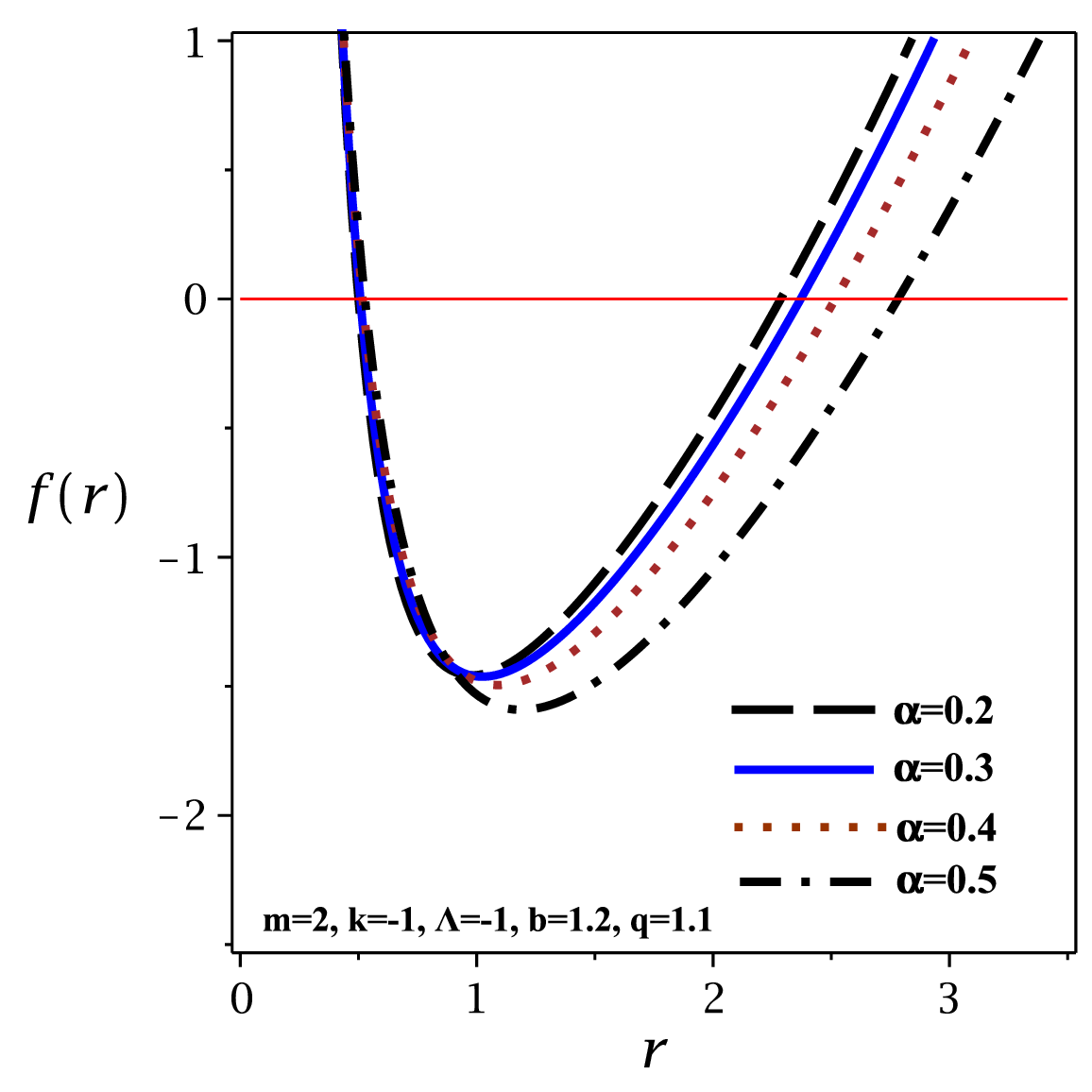} %
\includegraphics[width=0.35\textwidth]{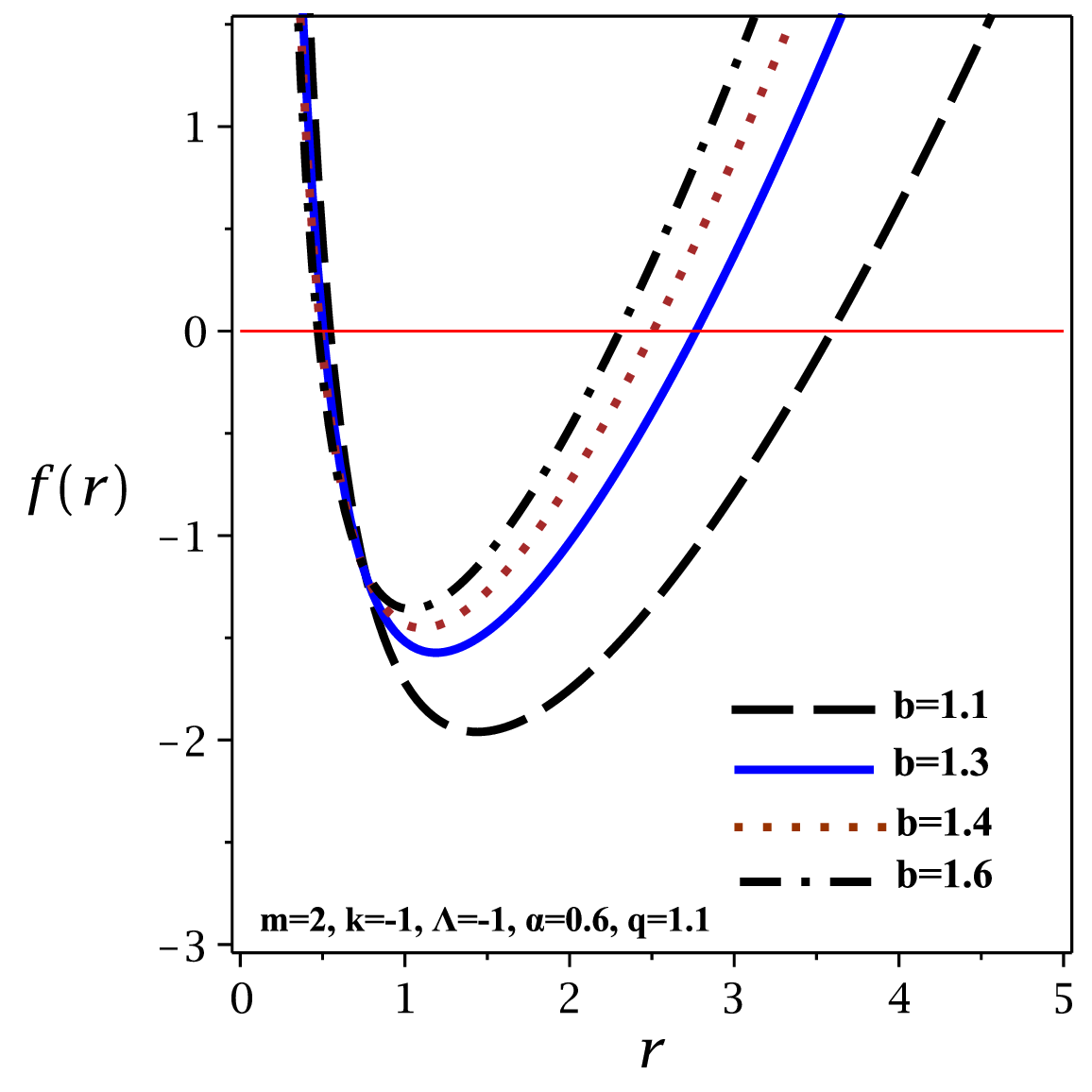} \newline
\caption{$f(r)$ versus $r$ for different values of $k$, $\protect\alpha$,
and $b$. The up panels are for $k=+1$. The middle panels are for $k=0$. The
down panels are for $k=-1$.}
\label{Fig2}
\end{figure}

Here, we plot the metric function $f(r)$ versus $r$ in Figs. \ref{Fig1} and %
\ref{Fig2} to demonstrate that the singularity at $r=0$ is indeed concealed
by an event horizon. Additionally, this visualization allows us to observe
the impact of dilaton gravity and the constant $k$ on the event horizon.

The effect of the constant $k$ on the root of the metric function is
illustrated in Fig. \ref{Fig1}. It is evident from the figure that there are
two roots. It is important to note that the outer root is associated with
the event horizon. Furthermore, our findings suggest that the larger black
holes correspond to $k=-1$, $k=0$, and $k=+1$, respectively, meaning that $%
r_{{+}_{k=-1}}>r_{{+}_{k=0}}>r_{{+}_{k=+1}}$ (where $r_{{+}_{k}}$ represents
the event horizon for different values of $k$).


Now, we are studying the effects of dilaton gravity ($\alpha$ and $b$) on
the event horizon of black holes for different values of $k$. Our findings
indicate the following:

1) The metric function ($f(r)$) plotted against $r$ for $k=+1$ is shown in
the up panels of Fig. \ref{Fig2}. The results show that as we increase $%
\alpha$, the root of the metric function decreases, and eventually there is
no root for the metric function (up left panel in Fig. \ref{Fig2}). In other
words, increasing $\alpha$ leads to the existence of a naked singularity.
However, there is a different behavior for $b$. In fact, as we increase $b$,
the root of the metric function increases, and we encounter large black
holes when $b$ is large.

2) For $k=0$, the middle panels of Fig. \ref{Fig2} show the plot of $f(r)$.
Similar to the previous case, the roots of the metric function increase
(decrease) as $b$ ($\alpha$) increases (decreases). Therefore, when we
consider small and large values of $\alpha$ and $b$ respectively, we find
large charged dilatonic black holes.

3) The metric function ($f(r)$) plotted against $r$ for $k=-1$ is shown in
the down panels of Fig. \ref{Fig2}. The results reveal that the behavior of
the root of the metric function is opposite to that of the two previous
cases. Indeed, for $k=-1$, large black holes have large values of $\alpha$
and small values of $b$.

\section{Thermodynamical Quantities}

Now, we can calculate the thermodynamic and conserved quantities of the
obtained solutions and determine whether these quantities satisfy the first
law of thermodynamics.

To analyze temperature, we employ the concept of surface gravity. The
temperature of these solutions is obtained as follows \cite%
{Sheykhi2007,HendiSPE2015} 
\begin{equation}
T=\frac{\mathcal{K}_{1,1}\left( \frac{b}{r_{+}}\right) ^{\frac{-2\alpha ^{2}%
}{\mathcal{K}_{1,1}}}}{4\pi }\left[ \frac{k}{\mathcal{K}_{-1,1}r_{+}}-\frac{%
q^{2}}{r_{+}^{3}}-\Lambda r_{+}\left( \frac{b}{r_{+}}\right) ^{\frac{4\alpha
^{2}}{\mathcal{K}_{1,1}}}\right] ,  \label{temp}
\end{equation}%
where $r_{+}$ is related to the event horizon and satisfies $f(r=r_{+})=0$.

The entropy of a black hole is determined using the area law. For
these black holes we get the entropy of black hole per unit volume in the
following form  
\begin{equation}
S=\frac{\widetilde{S}}{\mathcal{V}}=\frac{r_{+}^{2}}{4}\left( \frac{b}{r_{+}}%
\right) ^{\frac{2\alpha ^{2}}{\mathcal{K}_{1,1}}},  \label{entropy}
\end{equation}%
in which by setting $\alpha =0$, the entropy of dilatonic black holes
reduces to the entropy of black holes in Einstein-Maxwell's theory.

To determine the total electric charge of the solutions, Gauss's law can be
utilized. By calculating the flux of the electric field, we can determine 
the total electric charge per unit volume which is \cite%
{Sheykhi2007,HendiSPE2015} 
\begin{equation}
Q=\frac{\widetilde{Q}}{\mathcal{V}}=\frac{q}{4\pi }.  \label{Q}
\end{equation}

Next, our goal is to calculate the electric potential. We can achieve this
by analyzing the equation $F_{\mu \nu }=\partial _{\mu }A_{\nu }-\partial
_{\nu }A_{\mu }$. By doing so, we can identify the non-zero component of the
gauge potential, which is denoted as $A_{t}=-\int {F_{tr}dr}$. Consequently,
we can determine the electric potential at the event horizon ($U$) in
relation to the reference point ($r\rightarrow \infty $) as follows 
\begin{equation}
U\left( r\right) =-\int_{r_{+}}^{+\infty }F_{tr}dr=\frac{q}{r_{+}}.
\label{UU}
\end{equation}

Finally, the total mass per unit volume of the solution is
determined based on the definition of mass provided by Abbott and Deser \cite%
{AD1,AD2,AD3} (see Refs. \cite{Sheykhi2007,HendiSPE2015}, for more details) 
\begin{equation}
M=\frac{\widetilde{M}}{\mathcal{V}}=\frac{A}{8\pi \mathcal{K}_{-1,1}},
\label{Mass}
\end{equation}%
where 
\begin{equation}
A=\left( \frac{b}{r_{+}}\right) ^{\frac{4\alpha ^{2}}{\mathcal{K}_{1,1}}}%
\left[ kr_{+}-\frac{q^{2}\mathcal{K}_{1,-1}}{r_{+}}+\frac{\mathcal{K}_{1,1}%
\mathcal{K}_{-1,1}\Lambda r_{+}^{3}\left( \frac{b}{r_{+}}\right) ^{\frac{%
4\alpha ^{2}}{\mathcal{K}_{1,1}}}}{\mathcal{K}_{1,-3}}\right] .
\end{equation}

It is worth mentioning that when $\alpha = 0$, Eq. (\ref{Mass}) simplifies
to the mass of the Einstein-Maxwell black holes.

To determine if the quantities obtained, temperature (Eq. (\ref{temp})),
entropy (Eq. (\ref{entropy})), electric charge (Eq. (\ref{Q})), electric
potential (Eq. (\ref{UU})), and total mass of black holes (Eq. (\ref{Mass}),
satisfy the first law of thermodynamics, we show that 
\begin{equation}
\left( \frac{\partial M}{\partial S}\right) _{Q}=T\text{ \ \ \ \ }\&\text{\
\ \ \ \ }\left( \frac{\partial M}{\partial Q}\right) _{S}=U,
\end{equation}%
so, these quantities be able to satisfy the first law of thermodynamics,
which is 
\begin{equation}
dM=\left( \frac{\partial M}{\partial S}\right) _{Q}dS+\left( \frac{\partial M%
}{\partial Q}\right) _{S}dQ.
\end{equation}

It is notable that, the negative cosmological constant is treated as a
positive thermodynamical pressure in the extended phase space \cite%
{Lambdaext1,Lambdaext2}. In our context, it is expressed in the following
form \cite{HendiA2015} 
\begin{equation}
P=-\frac{\Lambda }{8\pi }\left( \frac{b}{r_{+}}\right) ^{\frac{2\alpha ^{2}}{%
\mathcal{K}_{1,1}}},  \label{P}
\end{equation}%
where is determined using the components of the energy-momentum tensor in
diagonal form. Moreover, in the absence of a dilaton field, the conventional
definition of pressure ($P=-\frac{\Lambda }{8\pi }$) holds true.

The Smarr formula is given by 
\begin{equation}
M=2\left( 1-\frac{\alpha ^{2}}{\mathcal{K}_{1,1}}\right) TS+UQ+2\left( \frac{%
2\alpha ^{2}}{\mathcal{K}_{1,1}}-1\right) VP,  \label{Smarr}
\end{equation}%
where in the absence of parameters of dilaton gravity ($\alpha =0$),
the Smarr formula turns to $M=2TS+UQ-2VP$.

The validity of the first law of thermodynamics in the extended
phase space should be written in the following form
\begin{equation}
dM=TdS+UdQ+VdP,  \label{FirstL}
\end{equation}%
where $V$ is the thermodynamical volume conjugate to $P$, and is
given by 
\begin{equation}
V=\left( \frac{\partial M}{\partial P}\right) =\left( \frac{\partial 
\widetilde{M}}{\mathcal{V}\partial P}\right) =\frac{\mathcal{K}_{1,1}}{%
\mathcal{K}_{-1,3}}b^{\frac{2\alpha ^{2}}{\mathcal{K}_{1,1}}}r_{+}^{\frac{%
\mathcal{K}_{1,3}}{\mathcal{K}_{1,1}}}.
\end{equation}%
where we use Eqs. (\ref{Mass}) and (\ref{P}) to extract the above equation.

Although it is theoretically possible to include additional terms such as $%
Bdb$ (where $B=\frac{\partial M}{\partial b}$) and $Ad\alpha $ (where $A=%
\frac{\partial M}{\partial \alpha }$) in the differential form of the first
law, doing so is not mathematically permissible. The Smarr formula (\ref%
{Smarr}) is expected to represent all intensive and extensive thermodynamic
parameters. Therefore, we treat $b$ and $\alpha $ as constants rather than
thermodynamic variables, as they do not appear in the Smarr formula.

\section{Thermodynamic Topology}

To explore thermodynamic topology, we employ the off-shell free energy
method, which views black hole solutions as topological defects within their
thermodynamic context. This approach involves analyzing both local and
global topology by calculating winding numbers associated with these
defects. These winding numbers categorize black holes based on their overall
topological charge. Importantly, a black hole's thermal stability correlates
with the sign of its winding number. The fundamental idea behind
thermodynamic topology revolves around understanding these topological
defects and their associated charges. Below, we discuss the essential
mathematical procedures involved in investigating thermodynamic topology.

In black hole thermodynamics, a vector field is derived from the generalized
off-shell free energy. The expression for the off-shell free energy of a
black hole with arbitrary mass, given by \cite{TheTo2} 
\begin{equation}
\mathcal{F}=E-\frac{S}{\tau },
\end{equation}%
where $E$ represents the energy (equivalent to the mass $M$) and $S$ denotes
the entropy of the black hole. The parameter $\tau$, representing the
timescale, is allowed to vary freely. To leverage this generalized free
energy, a vector field $\phi$ is defined as \cite{TheTo2} 
\begin{equation}
\phi=\left(\phi^r,\phi^\Theta \right)=\left(\frac{\partial\mathcal{F}}{%
\partial r_{+}},-\cot\Theta ~\csc\Theta \right).  \label{phifield}
\end{equation}

The zero points of this vector field are always obtained to be $%
(\tau,\Theta)=(\frac{1}{T},\frac{\pi}{2})$ where $T$ is the equilibrium
temperature of the cavity that encloses the black hole (on shell
temperature). At the points where a vector field either diverges or is not
defined, these locations carry significant physical implications.
Specifically, in our context, these points correspond to the zero points or
defects of the vector field, which represent black hole solutions. In
essence, we can interpret black holes as topological defects of the
constructed vector field. Consequently, each black hole solution possesses a
topological charge. We use Duan's $\phi$ mapping technique to determine the
associated topological charge. We calculate the unit vector $n$ of the field
in Eq. (\ref{phifield}), which are 
\begin{eqnarray}
n^{1} &=&\frac{\phi ^{r}}{\sqrt{(\phi ^{r})^{2}+(\phi ^{\Theta })^{2}}}, 
\notag \\
&&  \label{ns} \\
n^{2} &=&\frac{\phi ^{\Theta }}{\sqrt{(\phi ^{r})^{2}+(\phi ^{\Theta })^{2}}}%
,  \notag
\end{eqnarray}
The vector $n^a $ must satisfy two key conditions $n^a n^a=1$, and $n^a
\partial_\nu n^a=0$, where $\partial_\nu = \frac{\partial}{\partial x^\nu} $
and $\mu, \nu, \rho = 0, 1, 2 $.

We construct a topological current $j^{\mu }$ in the coordinate space $%
x^{\nu }=\{t,r_{+},\theta \}$ defined by 
\begin{equation}
j^{\mu }=\frac{1}{2\pi }\epsilon ^{\mu \nu \rho }\epsilon _{ab}\partial
_{\nu }n^{a}\partial _{\rho }n^{b},  \label{current}
\end{equation}%
the current $j^{\mu }$ is conserved, meaning $\partial _{\mu }j^{\mu }=0$.

The current in equation (\ref{current}) can be written as 
\begin{equation}
j^{\mu }=\delta ^{2}(\phi )J^{\mu }\left( \frac{\phi }{x}\right) ,
\label{current1}
\end{equation}%
where $J^{\mu }$ relates to the Jacobi tensor $\epsilon ^{ab}J^{\mu }\left( 
\frac{\phi }{x}\right) =\epsilon ^{\mu \nu \rho }\partial _{\nu }\phi
^{a}\partial _{\rho }\phi ^{b}$. We use the Laplacian Green function $%
\triangle _{\phi ^{a}}ln||\phi ||=2\pi \delta ^{2}(\phi )$ to get the
equation (\ref{current1}).

The topological charge $W$ relates to the zeroth component of the current
density as 
\begin{equation}
W=\int_{\Sigma }j^{0}d^{2}x=\sum_{i=1}^{N}w_{i},
\end{equation}%
where $w_{i}$ denotes the winding number around each zero point of the
vector field $\phi $. $\Sigma $ is the parameter region in which we
calculate the winding number $w_{i}$. Contours with appropriate dimensions
are constructed to define the parameter region, as follows 
\begin{equation}
\begin{cases}
r_{+}=r_{1}\cos \nu +r_{0}, \\ 
\\ 
\theta =r_{2}\sin \nu +\frac{\pi }{2},%
\end{cases}%
\end{equation}%
where $\nu \in (0,2\pi )$. Also, $r_{1}$ and $r_{2}$ are parameters that
define the dimensions of the contour to be formed and $r_{0}$ represents the
center point around which the contour is created. The winding numbers $w_{i}$
quantify how the vector field $n$ deflects along contours enclosing each
zero point. The mathematical relation between deflection angle $\Omega $ and
winding number is given by 
\begin{equation}
w_{i}=\frac{\Omega (2\pi )}{2\pi },  \label{winding}
\end{equation}%
where $\Omega $ is obtained by employing the contour integration $\Omega
(\nu )=\int_{0}^{\nu }\epsilon _{12}n^{1}\partial _{\nu }n^{2}d\nu$.

From Eq. (\ref{winding}), the topological charge is calculated by taking the
sum of the winding numbers calculated around all zero points of the vector
field $\phi$ or $W=\sum_{i=1}w_{i}$. This topological charge $W$ provides
insight into the structural properties of black hole solutions within the
framework of thermodynamic topology. It is important to highlight that $%
j^\mu $ is only non-zero at the zero points of the vector field $\phi$.
Thus, the total topological charge is considered zero if there are no zero
points within the parameter region.

In the next two subsections, we explore the thermodynamic topology of
topological charged dilatonic black holes. Our study focuses on two types of
thermodynamic setups: one where the charge ($q$) is fixed, and another where
the potential ($\phi$) is fixed. By calculating the topological properties
of these thermodynamic entities, we analyze both the local and global
structure of these black holes. We investigate how the topological constant (%
$k$) influences the thermodynamic characteristics. Additionally, we examine
how the parameters of dilaton gravity impact the thermodynamic
characteristics of these black holes, emphasizing differences from charged
black holes described by GR.

\subsection{\textbf{Fixed charge ($q$) ensemble}}

\subsubsection{\textbf{For elliptic ($k=1$) curvature hypersurface}}

In fixed charge ensemble, charge $q$ is a constant quantity. The off-shell
free energy is calculated using equations (\ref{entropy}) and (\ref{Mass}),
which leads to 
\begin{equation}
\mathcal{F}=M-\frac{S}{\tau }=\frac{\left( \frac{b}{r_{+}}\right) {}^{\frac{%
4\alpha ^{2}}{\mathcal{K}_{1,1}}}\left( \frac{\mathcal{K}_{1,1}\Lambda
r_{+}^{4}\left( \frac{b}{r_{+}}\right) {}^{\frac{4\alpha ^{2}}{\mathcal{K}%
_{1,1}}}}{\mathcal{K}_{1,-3}}-\frac{kr_{+}^{2}}{\mathcal{K}_{1,-1}}%
+q^{2}\right) }{8\pi r_{+}}-\frac{r_{+}^{2}}{4\tau }\left( \frac{b}{r_{+}}%
\right) {}^{\frac{2\alpha ^{2}}{\mathcal{K}_{1,1}}}.
\end{equation}

Taking the first order derivative of the off-shell free energy, the $\phi
^{r}$ component is found to be 
\begin{equation}
\phi ^{r}=\frac{\left( \frac{b}{r_{+}}\right) ^{\frac{4\alpha ^{2}}{\mathcal{%
K}_{1,1}}}}{8\pi \mathcal{K}_{1,1}r_{+}^{2}}\left( \frac{\mathcal{K}%
_{3,-1}kr_{+}^{2}}{\mathcal{K}_{1,-1}}-\frac{\mathcal{K}_{1,1}\mathcal{K}%
_{5,-3}\Lambda r_{+}^{4}\left( \frac{b}{r_{+}}\right) ^{\frac{4\alpha ^{2}}{%
\mathcal{K}_{1,1}}}}{\mathcal{K}_{1,-3}}-\frac{4\pi r_{+}^{3}}{\tau \left( 
\frac{b}{r_{+}}\right) ^{\frac{2\alpha ^{2}}{\mathcal{K}_{1,1}}}}-q^{2}%
\mathcal{K}_{5,1}\right) .
\end{equation}

Solving $\phi ^{r}=0$, $\tau $ can be obtained in the following form 
\begin{equation}
\tau =\frac{4\pi r_{+}^{3}\left( \frac{b}{r_{+}}\right) ^{\frac{-2\alpha ^{2}%
}{\mathcal{K}_{1,1}}}}{\frac{\mathcal{K}_{3,-1}kr_{+}^{2}}{\mathcal{K}_{1,-1}%
}-\frac{\mathcal{K}_{1,1}\mathcal{K}_{5,-3}\Lambda r_{+}^{4}\left( \frac{b}{%
r_{+}}\right) {}^{\frac{4\alpha ^{2}}{\mathcal{K}_{1,1}}}}{\mathcal{K}_{1,-3}%
}-q^{2}\mathcal{K}_{5,1}}.  \label{tau}
\end{equation}%
\begin{figure}[h]
\centering
\begin{subfigure}{0.28\textwidth}
		\includegraphics[width=5cm,height=5cm]{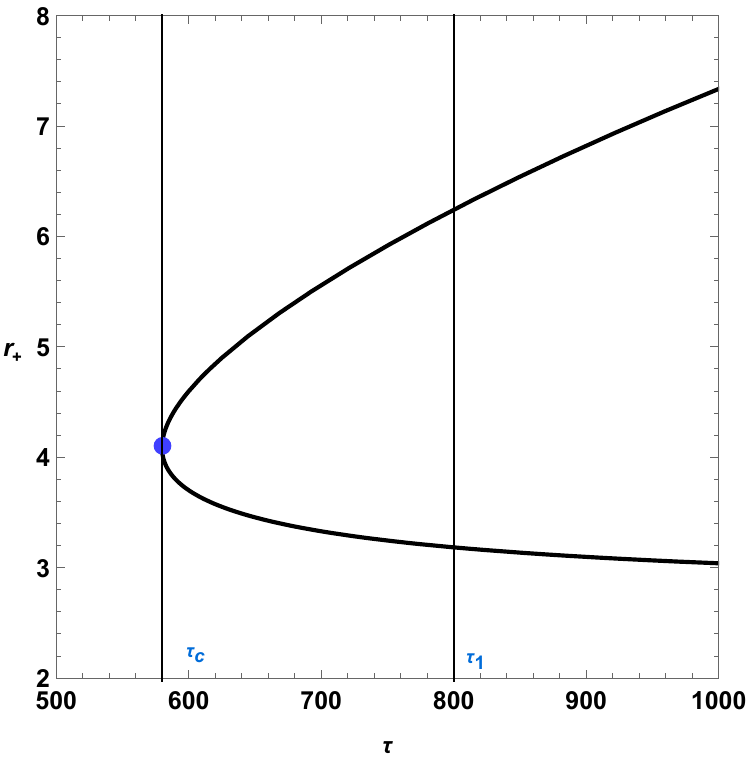}
		\caption{}
		\label{t1a}
	\end{subfigure}
\begin{subfigure}{0.27\textwidth}
		\includegraphics[height=5cm,width=6cm]{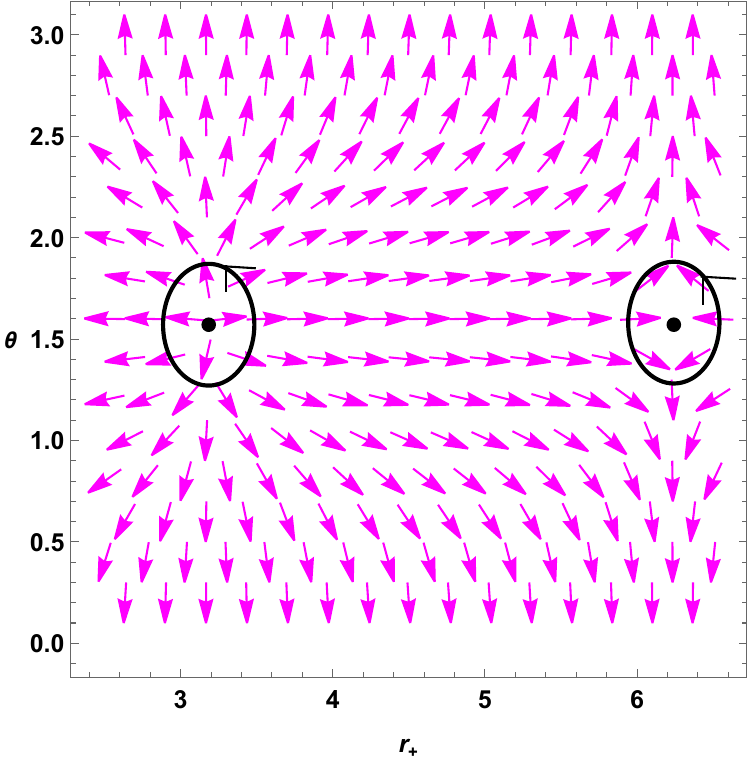}
		\caption{}
		\label{t1b}
	\end{subfigure}\hspace{0.6cm} 
\begin{subfigure}{0.38\textwidth}
		\includegraphics[width=5cm,height=5cm]{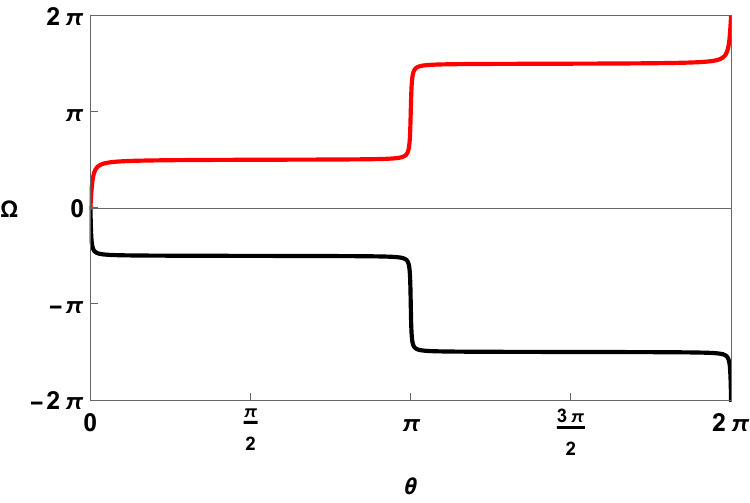}
		\caption{}
		\label{t1c}
	\end{subfigure}
\caption{Topological charge of topological charged dilatonic black holes $%
(k=+1)$ in fixed charge ensemble when positive values of $\Lambda $ is
considered. Here we have considered $q=1$, $\protect\alpha =0.5$, $b=1$, $%
k=+1 $, $\Lambda =0.01$.}
\label{c1}
\end{figure}

Next, $\tau $ vs $r_{+}$ is plotted in Fig. \ref{t1a} for $q=1$, $\alpha
=0.5 $, $b=1$, $\Lambda =0.01$ and $k=+1$. This plot reveals two black hole
branches: a small black hole branch for the range $r_{+}<4.0867$ and a large
black hole branch for $r_{+}>4.0867$. To calculate topological charge of the
black hole we choose a random value of $\tau =\tau_{1}=800$ and calculate
the zero point of the vector field. In Fig. \ref{t1b}, the vectors represent
a part of the ($n^{1}$,$~n^{2}$) field in the $r_{+}-\theta $ plane, where
the zero points for $\tau =\tau_{1}=800$ are observed at $r_{+}=3.1842$ and $%
r_{+}=6.24193$. The topological charge is calculated in Fig. \ref{t1c},
where the winding number corresponding to $r_{+}=3.1842$ is $+1$ represented
by the red solid line and the winding number corresponding to $r_{+}=6.24193$
is $-1$, represented by the black solid line. By adding the winding number
the topological charge $W$ is obtained as $W=1-1=0.$ We solve the equation $%
\frac{\partial \tau}{\partial r_+}=0$ to obtain the exact point at which the
phase transition takes place i.e., $(\tau _{c}$, $r_{c})=(579.841$, $4.0867)$
represented by the blue dot in Fig. \ref{t1a}. It is important to mention
that positive winding number represent a stable black hole branch and
negative winding number represents the opposite.In Fig. \ref{t1a},the small
black hole branch is the stable branch and large black hole branch is the
unstable branch.Hence the phase transitioning point $(\tau _{c}$, $r_{c})$
is called an annihilation point as we make a transition from a stable black
hole branch to an unstable black hole branch.

\begin{figure}[h!]
\centering
\begin{subfigure}{0.28\textwidth}
		\includegraphics[width=\linewidth]{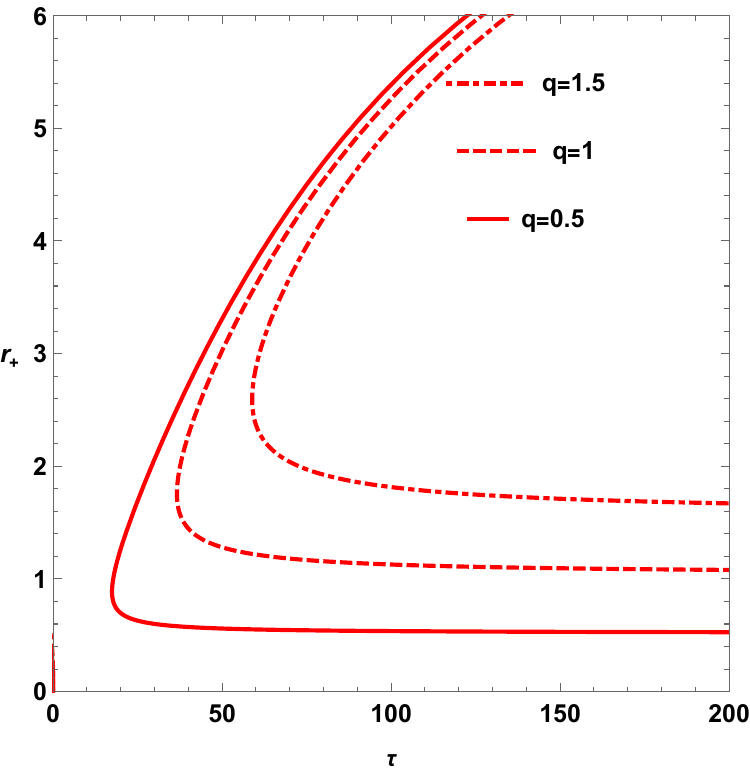}
		\caption{$\alpha=0.1$, $b=1$, $k=1$, and $\Lambda=0.01$.}
		\label{t2a}
	\end{subfigure}
\begin{subfigure}{0.28\textwidth}
		\includegraphics[width=\linewidth]{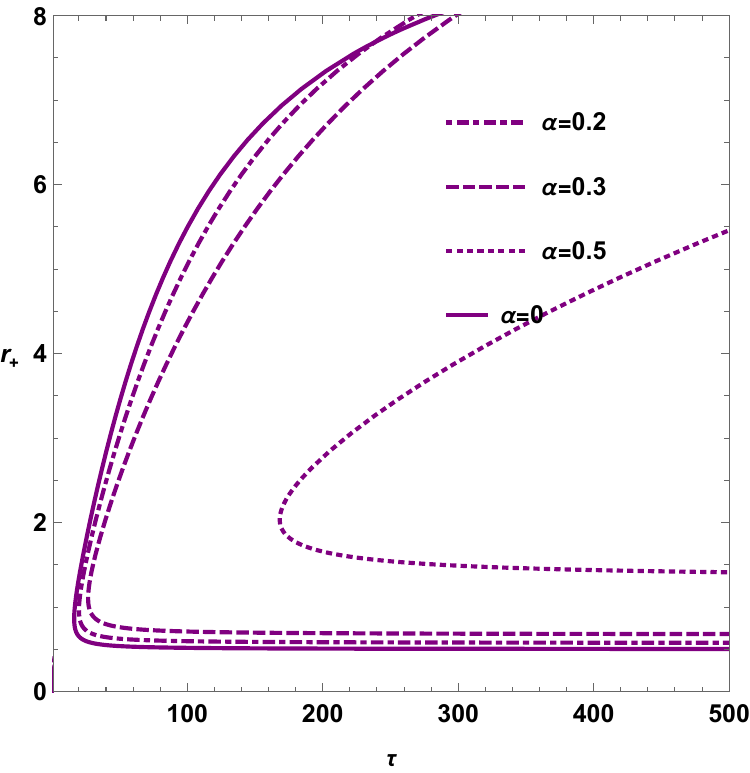}
		\caption{$q=0.5$, $b=1$, $k=1$, and $\Lambda=0.01$.}
		\label{t2b}
	\end{subfigure} \hspace{0.1cm} 
\begin{subfigure}{0.28\textwidth}
		\includegraphics[width=\linewidth]{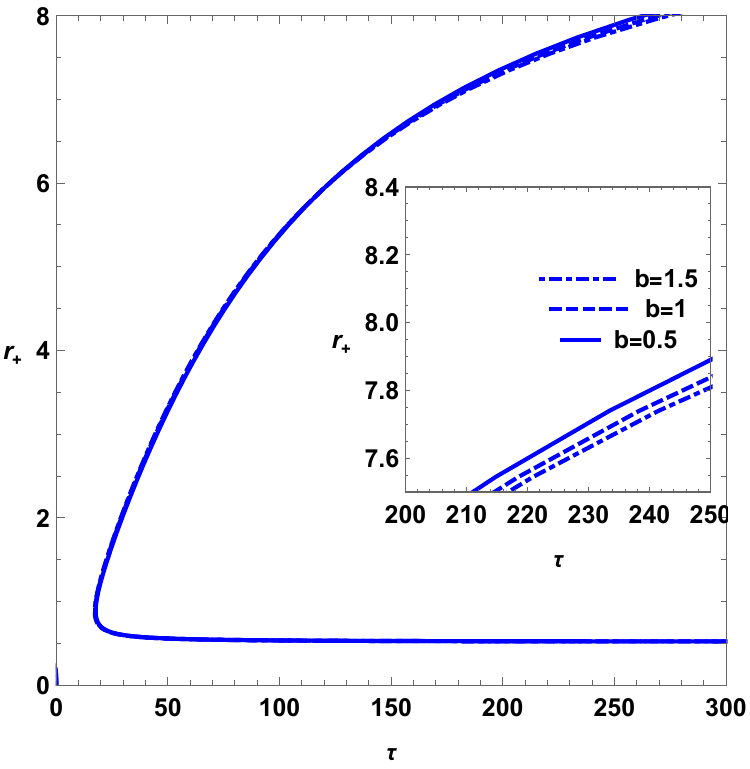}
		\caption{$q=0.5$, $\alpha=0.1$, $k=1$, and $\Lambda= 0.01$.}
		\label{t2c}
	\end{subfigure}
\caption{ Variation of $\protect\tau $ vs $r_{+}$ plots for topological
charged dilaton black holes $(k=+1)$ with different thermodynamic
parameters. }
\label{c2}
\end{figure}

Moreover, the topological charge remains invariant with variation of
thermodynamic parameters as illustrated in Figure. \ref{c2}, the topological
charge is found to be $0$ across all cases.

\begin{figure}[t!]
\centering
\begin{subfigure}{0.33\textwidth}
		\includegraphics[width=\linewidth]{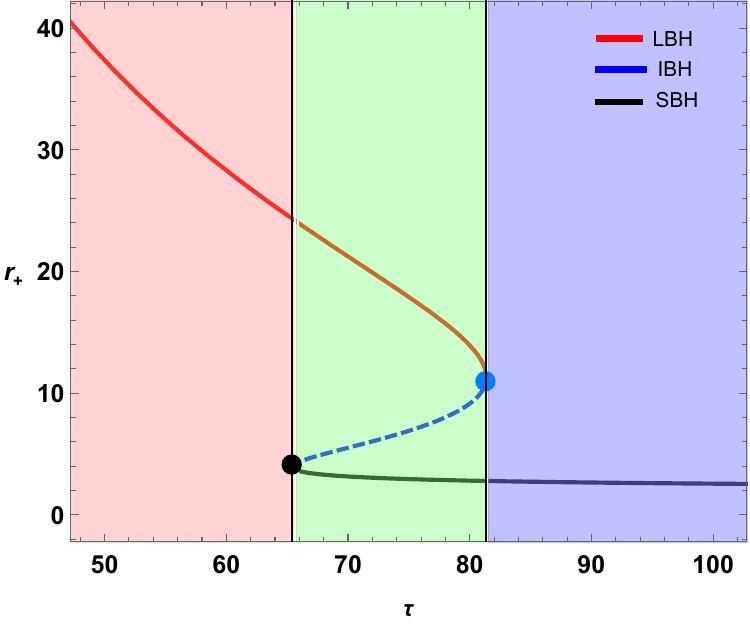}
		\caption{}
		\label{c3a}
	\end{subfigure} 
\begin{subfigure}{0.28\textwidth}
		\includegraphics[width=\linewidth]{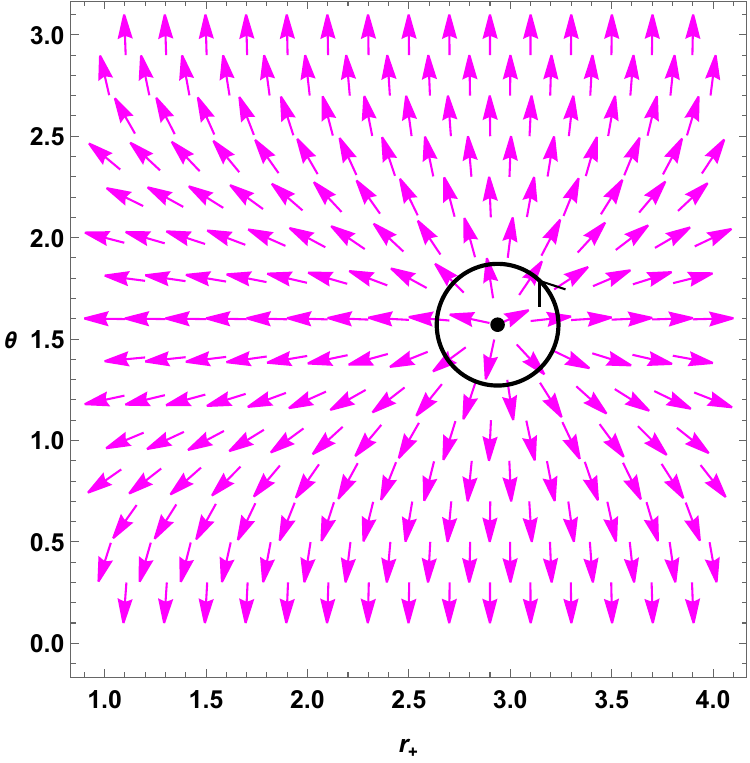}
		\caption{}
		\label{c3b}
	\end{subfigure}
\begin{subfigure}{0.28\textwidth}
		\includegraphics[width=\linewidth]{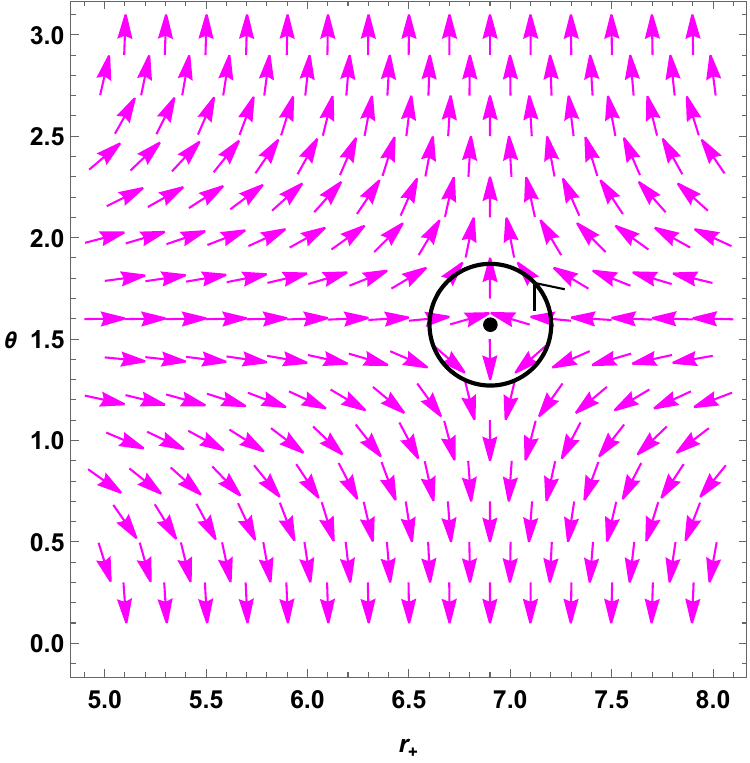}
		\caption{}
		\label{c3c}
	\end{subfigure}
\begin{subfigure}{0.28\textwidth}
		\includegraphics[width=\linewidth]{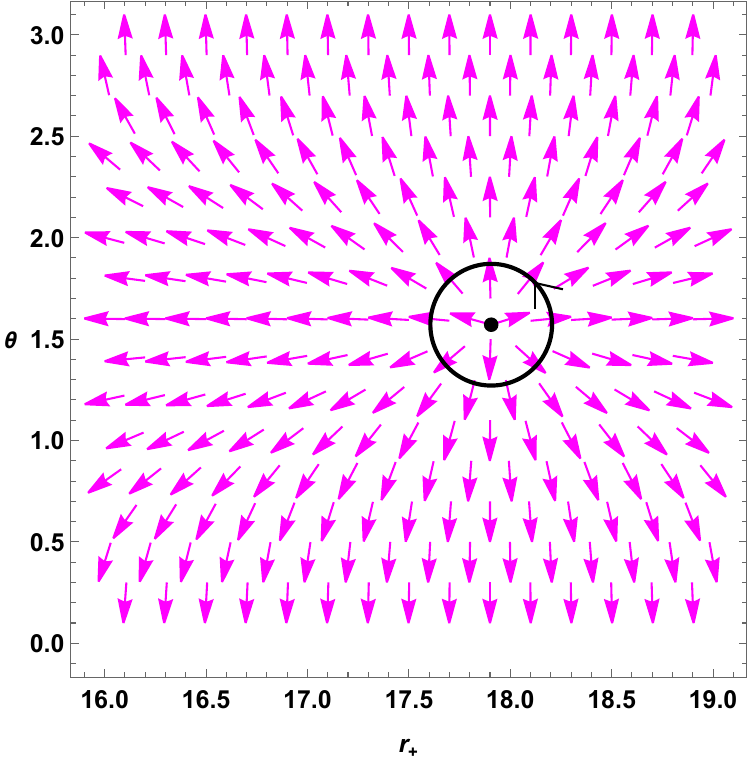}
		\caption{}
		\label{c3d}
	\end{subfigure} \hspace{0.8cm} 
\begin{subfigure}{0.37\textwidth}
	\includegraphics[width=\linewidth]{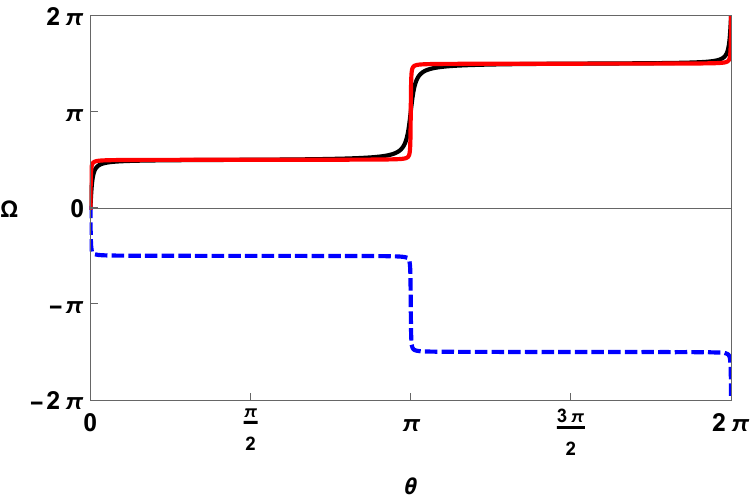}
	\caption{}
	\label{c3e}
\end{subfigure}
\hspace{1.5cm}
\caption{Fig. \protect\ref{c3a}, represents $\protect\tau$ vs $r_+$ plot for 
$q=2$, $b=1$, $\Lambda=-0.01$, $k=+1$, and $\protect\alpha=0.15$. Fig. 
\protect\ref{c3b}, Fig. \protect\ref{c3c}, and Fig. \protect\ref{c3d} shows
vector plot for zero points $r_{+}=2.9375$, $r_{+}=6.2419$, $r_{+}=17.9060$,
respectively. The winding number calculations for zero points $r_{+}=2.9375$%
, $r_{+}=6.2419$, $r_{+}=17.9060$ are represented by black solid line, blue
dashed line, red solid line respectively in Fig. \protect\ref{c3e}.}
\label{c3}
\end{figure}

\begin{figure}[tbp]
\begin{subfigure}{0.28\textwidth}
		\includegraphics[width=\linewidth]{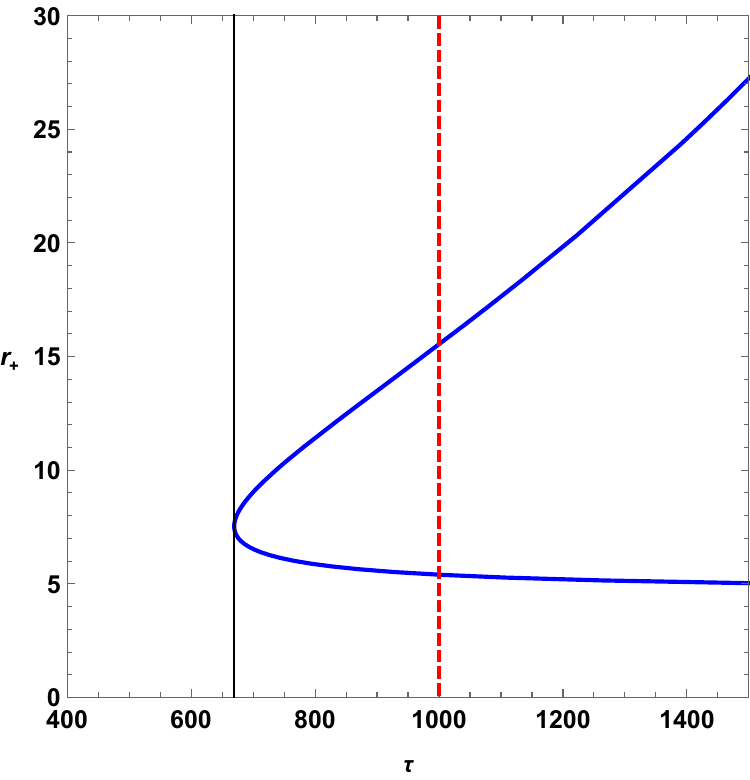}
		\caption{}
		\label{c4a}
	\end{subfigure}
\begin{subfigure}{0.28\textwidth}
		\includegraphics[width=\linewidth]{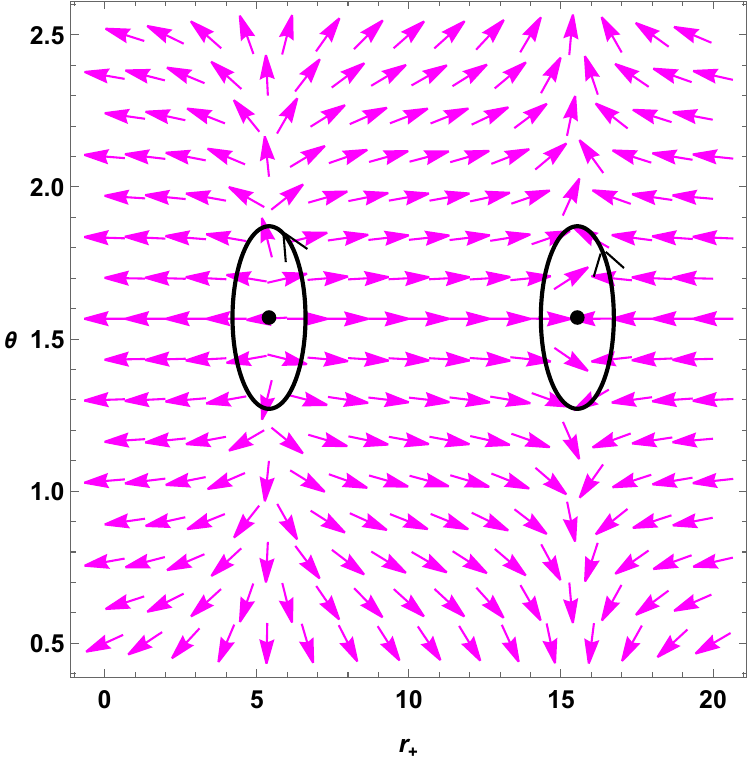}
		\caption{}
		\label{c4b}
	\end{subfigure}
\begin{subfigure}{0.34\textwidth}
		\includegraphics[width=\linewidth]{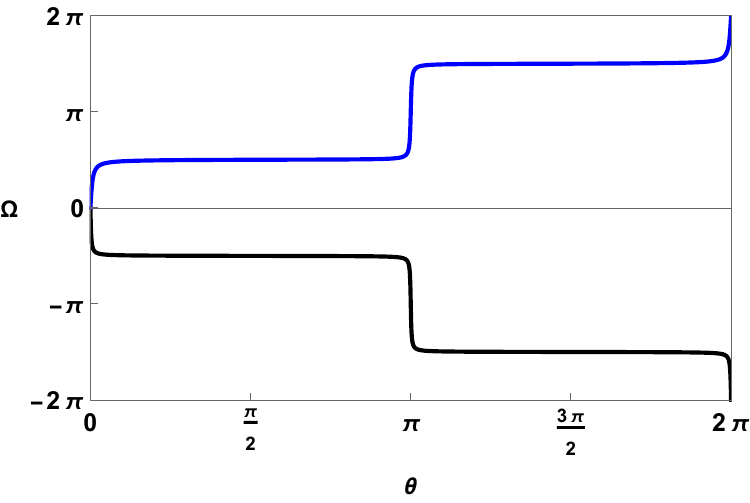}
		\caption{}
		\label{c4c}
	\end{subfigure}
\caption{Fig. \protect\ref{c4a} is the $\protect\tau$ vs $r_+$ plot for $%
\protect\alpha=0.49$ and $q=2$, $b=1$, $\Lambda=-0.01$, $k=+1$. Fig. \protect
\ref{c4b} is the vector plot in the $r-\protect\theta$ plane for $\protect%
\tau=1000$. As it is clear from the vector plot, the zero point is found to
be at $(5.4030, \frac{\protect\pi}{2})$ in the small black hole branch and $%
(15.5381, \frac{\protect\pi}{2})$ in the large black hole branch. In Fig. 
\protect\ref{c4c}, the blue contour represents winding number for large
black hole branch and black contour represents the same for small black hole
branch.}
\label{c4}
\end{figure}

When considering a negative value of $\Lambda $, we observe that these black
holes can be categorized into two topological classes with topological
charges of either $0$ or $1$, depending on the value of $\alpha $.
Interestingly, the phase transition properties vary with changes in $\alpha $%
. For smaller values of $\alpha $ (close to zero), these black holes exhibit
van der Waals-like phase transition. At $\alpha =0$, the black holes mimic
the characteristics of Reissner-Nordstrom (RN) charged AdS black holes. As
the value of $\alpha $ increases, only the Davies-type phase transition is
observed. Consequently, the thermodynamic topology of these black holes
changes with the increasing value of $\alpha $. In Fig. \ref{c3} and Fig. %
\ref{c4}, we plot $\tau $ vs $r_{+}$ for two values of $\alpha$ for example,
while keeping $q=2$, $b=1$, $\Lambda =-0.01$, and $k=1$. In Fig. \ref{c3a}
for $\alpha =0.15$, we can clearly see three black hole branches. Vector
plots in Fig. \ref{c3b}, Fig. \ref{c3c} and Fig. \ref{c3d}, represent the
zero point in the $(r,\theta )$ plane for $\tau =75$. There are three zero
points: $(2.9375,\frac{\pi }{2})$ in the small black hole branch (SBH), $%
(6.2419,\frac{\pi }{2})$ in the intermediate black hole branch (IBH) and $%
(17.9060,\frac{\pi }{2})$ in the large black hole branch (LBH). Next, we
will calculate the winding number for all three branches at the zero points
mentioned above and by adding them we will obtain the total topological
charge. It is found that the small black hole (SBH) branch, depicted by the
black solid line in the range $0\leq r_{+}\leq3.9272 $, has a winding number
of $+1$. The intermediate black hole branch, shown by the blue dashed line
for $3.9272<r_{+}\leq 11.3159$, has a winding number of $-1$. The large
black hole branch, represented by the red solid line for $r_{+}>11.3159$,
also has a winding number of $+1$. These winding numbers are illustrated in
Fig. \ref{c3e}. The winding numbers indicate that both the large and small
black hole branches are stable (with a winding number of $+1$), while the
intermediate black hole branch is unstable (with a winding number of $-1$).
Consequently, the total topological charge is $W=1-1+1=1$. Moreover, for
this value of $\alpha $, we observe an annihilation point at $r_{+}=3.9272$
and a generation point at $r_{+}=11.3159 $. Then, we consider $\alpha =0.49$
and keep the rest of the value the same as the above scenario.

In Fig. \ref{c4a}, we observe two black hole branches: a stable small black
hole branch with winding number $+1$ and an unstable large black hole branch
with winding number $-1$. An annihilation point is observed at $r_{+}=7.5278$
The total topological charge is found to be $W=-1+1=0.$. Thus, it can be
inferred that the topological charge changes with variations in the value of 
$\alpha$. However, other thermodynamic parameters are found to have no
impact on the topological charge.

\begin{figure}[tbp]
{b!} 
\begin{subfigure}{0.28\textwidth}
		\includegraphics[width=\linewidth]{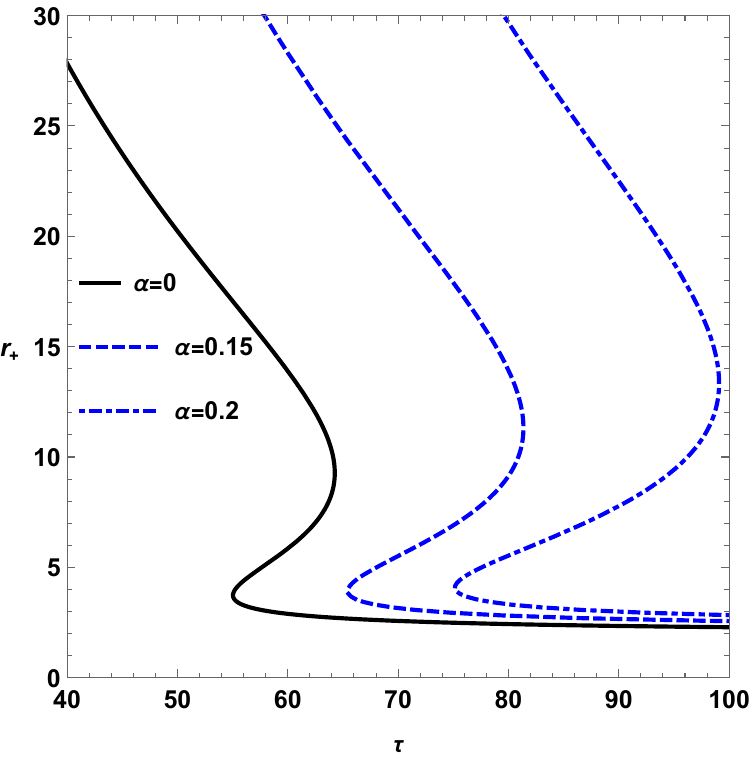}
		\caption{W=1}
		\label{c5a}
	\end{subfigure}
\hspace{0.5cm} 
\begin{subfigure}{0.28\textwidth}
		\includegraphics[width=\linewidth]{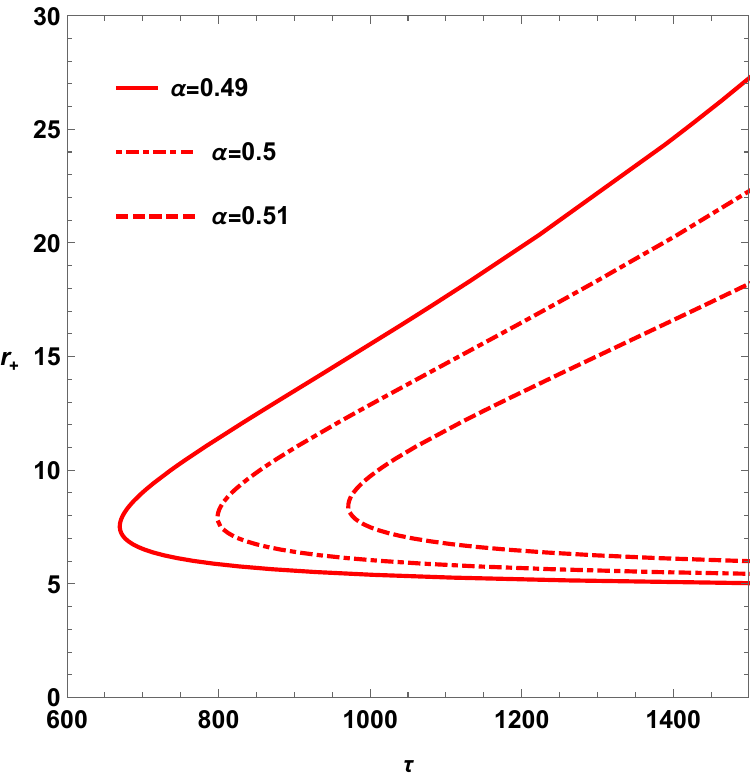}
		\caption{W=0}
		\label{c5b}
	\end{subfigure}
\caption{$\protect\tau$ vs $r_+$ plot for different values $\protect\alpha$
while keeping $q=2$, $b=1$, $\Lambda=-0.01$, and $k=+1$ fixed. $W$ denotes
the winding number for the plots.}
\label{c55}
\end{figure}

In Fig. \ref{c55}, the impact of $\alpha$ on topological charge is shown. It
is observed that, apart from $\alpha$ and sign of $\Lambda$, the topological
charge is independent of other thermodynamic parameters. In conclusion, for
topological charged dilatonic black holes with $k=+1$ in the fixed charge
ensemble, the topological charge is $0$ for positive value of $\Lambda$ and $%
0$ or $1$ for the negative value of $\Lambda$ depending on the value of $%
\alpha$. The critical value of $\alpha$(we denote as $\alpha_c$) at
which the topological changes from $W=1$ to $W=0$ is dependent on $b$. The
exact value at which this transition takes place can be found by solving the
equation for $\alpha_c$ 
\begin{equation*}
\biggl(\frac{\partial \tau}{\partial r_+}\biggr)_{\alpha=\alpha_c}=0,
\end{equation*}
As the equation is very complicated to obtain an analytical relation between 
$b$ and $\alpha_c$, we chose to fix the value of $b$ and other thermodynamic
quantities to get a numerical value of $\alpha_c$. For example, in Fig. 5 of
the manuscript, we plot $\tau$ vs $r_{+}$ for $\alpha=0.15$ while keeping $%
q=2$, $b=1$, $\Lambda=-0.01$, and $k=1$, where we can clearly see three
black hole branches. The topological charge in that case is found to be $W=1$%
, and the critical value of $\alpha$ is found to be $\alpha_c=0.4473$. Above 
$\alpha_c$, we found topological charge $W=0$. In Table \ref{tabr}, the
variation of $\alpha_c$ is shown.

\begin{table}[ht!]
\caption{Variation of critical value of $\protect\alpha_c$ with $b$ while
keeping $q=2$ and $\Lambda=-0.01$ fixed}
\label{tabr}\centering
\begin{adjustbox}{max width=\textwidth}
\begin{tabular}{|c|c|c|c|}
\hline
\textbf{Value of $b$} & \textbf{Value of $\alpha_c$} & \textbf{Value of $b$} & \textbf{Value of $\alpha_c$} \\
\hline
0.1  & 0.4502 & 0.8  & 0.4428 \\
0.2  & 0.4497 & 0.9  & 0.4464 \\
0.3  & 0.4407 & 1.0  & 0.4473 \\
0.35 & 0.4569 & 1.1  & 0.4473 \\ 
0.4  & 0.4212 & 1.2  & 0.4473 \\
0.5  & 0.4282 & 1.3  & 0.4473 \\
0.6  & 0.4393 & 1.4  & 0.4473 \\
0.7  & 0.4387 & 1.5  & 0.4473 \\ 
\hline
\end{tabular}
\end{adjustbox}
\end{table}

\vspace{2cm}

\subsubsection{\textbf{For hyperbolic ($k=-1$) curvature hypersurface}}

\begin{figure}[h]
\centering
\begin{subfigure}{0.28\textwidth}
		\includegraphics[width=\linewidth]{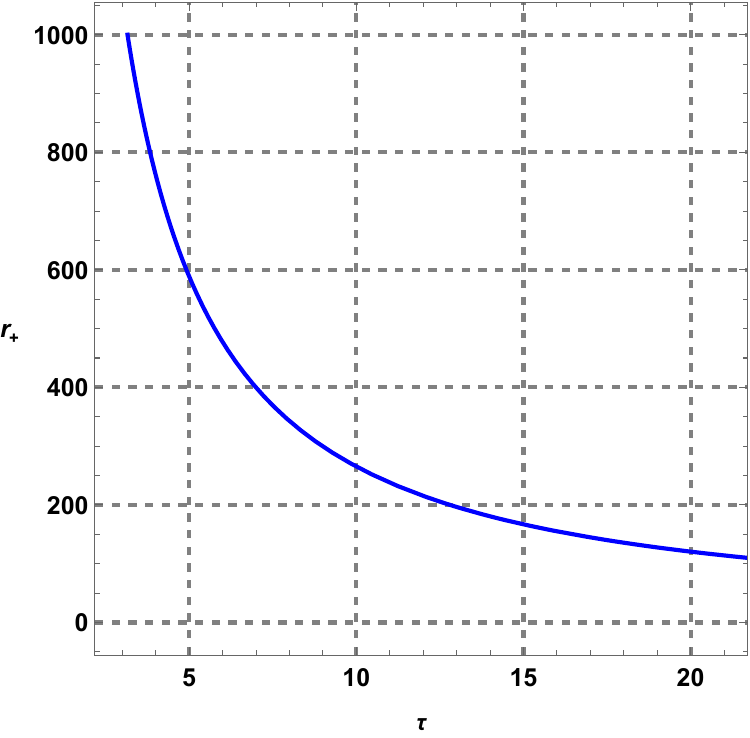}
		\caption{}
		\label{c6a}
	\end{subfigure}
\begin{subfigure}{0.27\textwidth}
		\includegraphics[width=\linewidth]{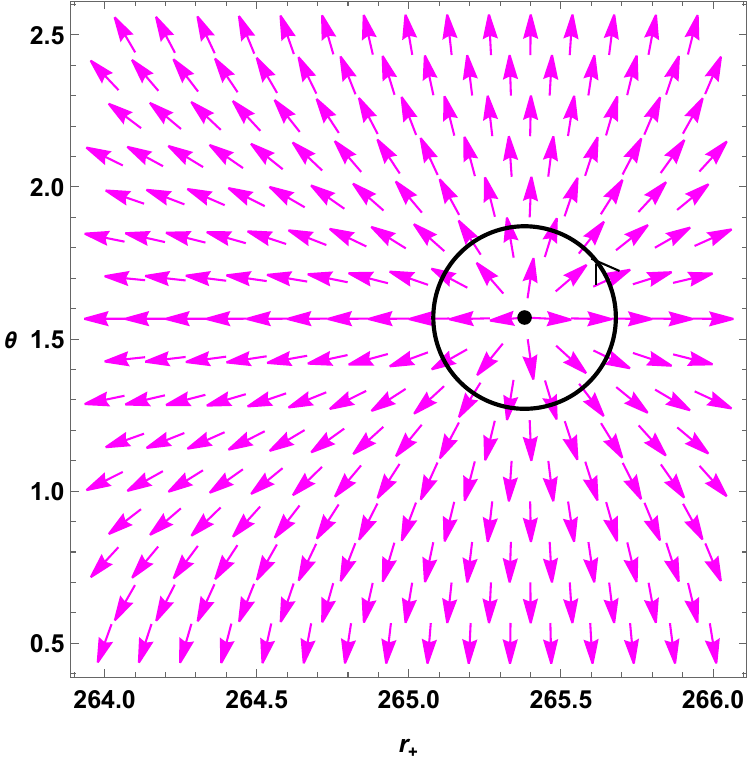}
		\caption{}
		\label{c6b}
	\end{subfigure} \hspace{0.2cm} 
\begin{subfigure}{0.35\textwidth}
		\includegraphics[width=\linewidth]{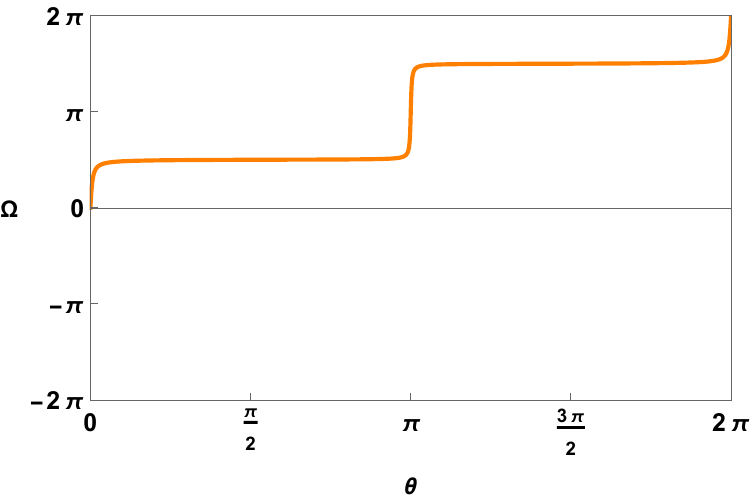}
		\caption{}
		\label{c6c}
	\end{subfigure}
\caption{Fig. \protect\ref{c6a} is the $\protect\tau$ vs $r_+$ plot for $%
\protect\alpha=0.15$ and $q=2$, $b=1$, $\Lambda=-0.01$, $k=-1$. Fig. \protect
\ref{c6b} is the vector plot in the $r_{+}-\protect\theta$ plane for $%
\protect\tau=10$. As it is clear from the vector plot, the zero point is
found to be at $(265.3804, \frac{\protect\pi}{2})$. In Fig. \protect\ref{c6c}%
, the orange contour represents the winding number for the black hole, which
equals $1$.}
\label{c6}
\end{figure}

\begin{figure}[h!]
\centering
\begin{subfigure}{0.29\textwidth}
		\includegraphics[width=\linewidth]{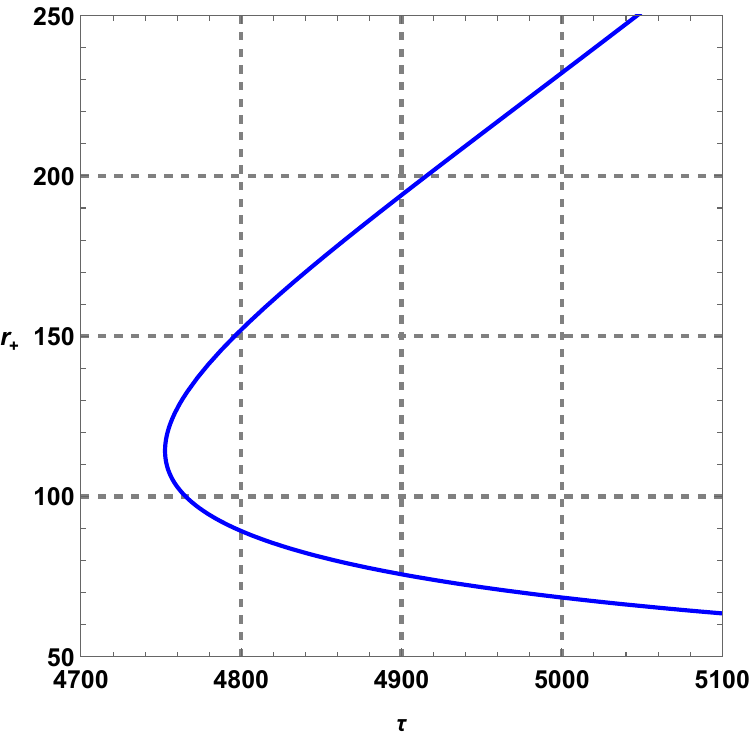}
		\caption{}
		\label{c7a}
	\end{subfigure}
\hspace{0.5cm} 
\begin{subfigure}{0.28\textwidth}
		\includegraphics[width=\linewidth]{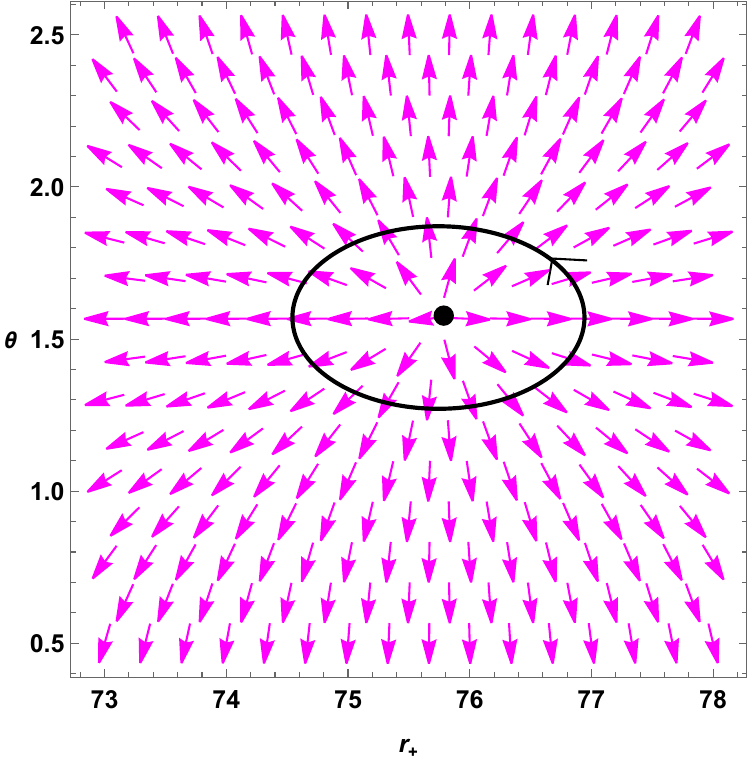}
		\caption{}
		\label{c7b}
	\end{subfigure}\newline
\begin{subfigure}{0.28\textwidth}
	\includegraphics[width=\linewidth]{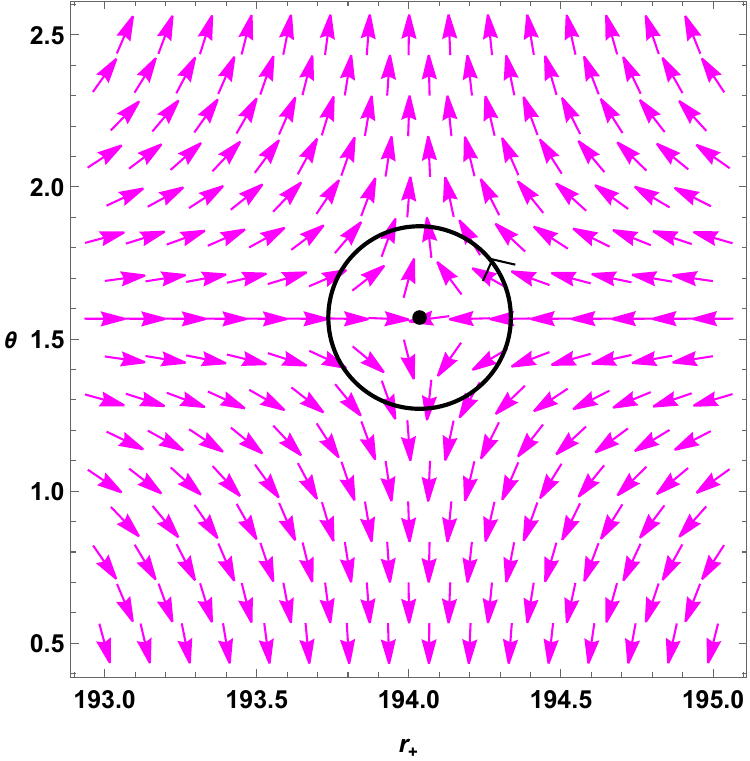}
	\caption{}
	\label{c7c}
\end{subfigure}
\hspace{0.5cm} 
\begin{subfigure}{0.35\textwidth}
		\includegraphics[width=\linewidth]{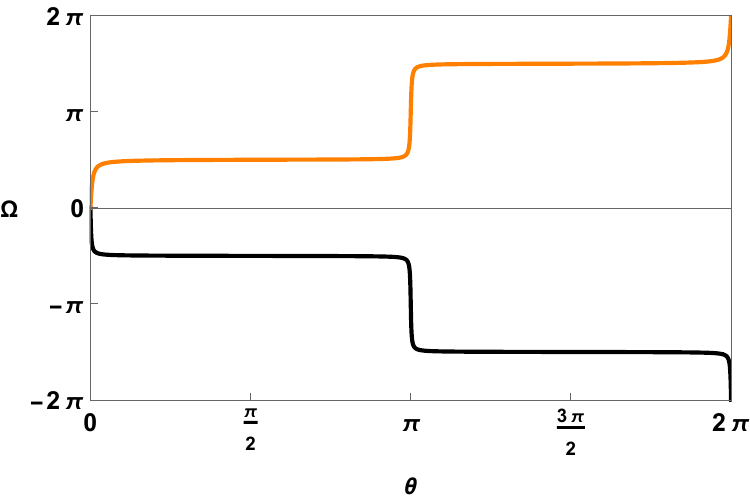}
		\caption{}
		\label{c7d}
	\end{subfigure}
\caption{Fig. \protect\ref{c7a} is the $\protect\tau$ vs $r_+$ plot for $%
\protect\alpha=0.5$ and $q=2$, $b=1$, $\Lambda=-0.01$, $k=-1$. Fig. \protect
\ref{c7b} and Fig. \protect\ref{c7c} are the vector plot in the $r_{+}-%
\protect\theta$ plane for $\protect\tau=4900$. As it is clear from the
vector plots, the zero points are found to be at $(75.7431, \frac{\protect\pi%
}{2})$ in the small black hole branch and $(194.0352, \frac{\protect\pi}{2})$
in the large black hole branch. In Fig. \protect\ref{c7d}, the orange
contour represents the winding number for the small black hole branch and
the black contour represents the same for the large black hole branch.}
\label{c7}
\end{figure}

For topological charged dilatonic black holes with constant $t$ and $r$
boundaries, characterized by a hypersurface with hyperbolic curvature, the
equation for $\tau $ can be derived by substituting $k=-1$ into Eq. (\ref%
{tau}), which is

\begin{equation}
\tau _{k=-1}=\frac{4\pi r_{+}^{3}\left( \frac{b}{r_{+}}\right) ^{\frac{%
-2\alpha ^{2}}{\mathcal{K}_{1,1}}}}{\frac{\mathcal{K}_{3,-1}r_{+}^{2}}{%
\mathcal{K}_{-1,1}}-\frac{\mathcal{K}_{1,1}\mathcal{K}_{5,-3}\Lambda
r_{+}^{4}\left( \frac{b}{r_{+}}\right) {}^{\frac{4\alpha ^{2}}{\mathcal{K}%
_{1,1}}}}{\mathcal{K}_{1,-3}}-q^{2}\mathcal{K}_{5,1}},  \label{taukN1}
\end{equation}%
From the above equation, it is clear that, because of the requirement for a
positive temperature, $\Lambda $ must be negative. We found two topological
classes for these kinds of black holes. One with topological charge $+1$
which has a single branch in $\tau $ vs $r_{+}$ plot and another topological
class with charge $0$ which contains one stable small black hole branch
(winding number$=+1$) and an unstable black hole branch (winding number$=-1$%
). Figs. \ref{c6} and \ref{c7} demonstrate the different topological class
exhibited by these black holes.



\subsubsection{\textbf{For flat ($k=0$) curvature hypersurface}}

Substituting $k=0$ in Eq. (\ref{tau}), the expression of $\tau $, for
topological charged dilatonic black holes with the boundary of $t=$ constant
and $r=$constant becomes 
\begin{equation}
\tau _{k=0}=\frac{4\pi r_{+}^{3}\left( \frac{b}{r_{+}}\right) ^{\frac{%
-2\alpha ^{2}}{\mathcal{K}_{1,1}}}}{-\frac{\mathcal{K}_{1,1}\mathcal{K}%
_{5,-3}\Lambda r_{+}^{4}\left( \frac{b}{r_{+}}\right) {}^{\frac{4\alpha ^{2}%
}{\mathcal{K}_{1,1}}}}{\mathcal{K}_{1,-3}}-q^{2}\mathcal{K}_{5,1}}.
\label{tauk0}
\end{equation}

\begin{figure}[h]
\centering
\begin{subfigure}{0.28\textwidth}
		\includegraphics[width=\linewidth]{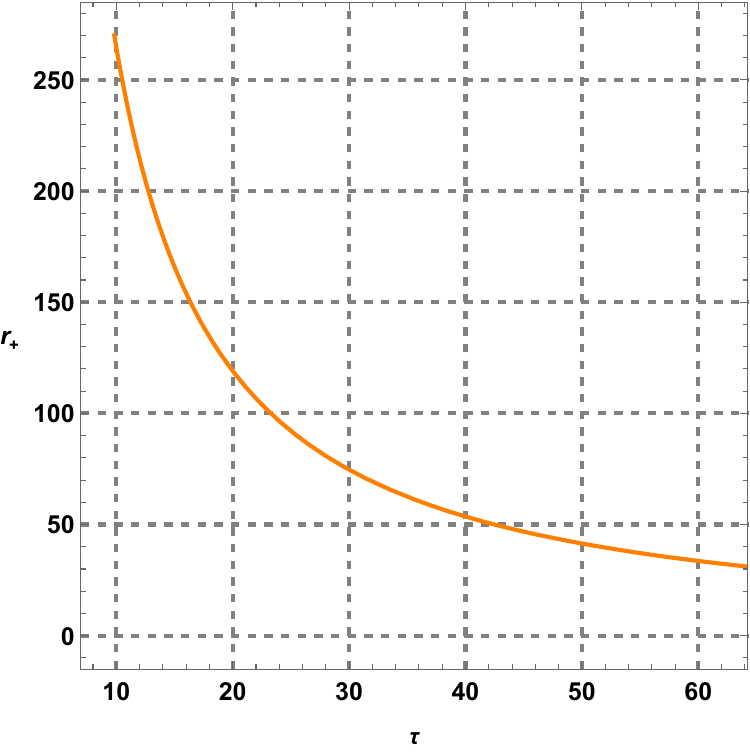}
		\caption{}
		\label{c8a}
	\end{subfigure}
\begin{subfigure}{0.28\textwidth}
		\includegraphics[width=\linewidth]{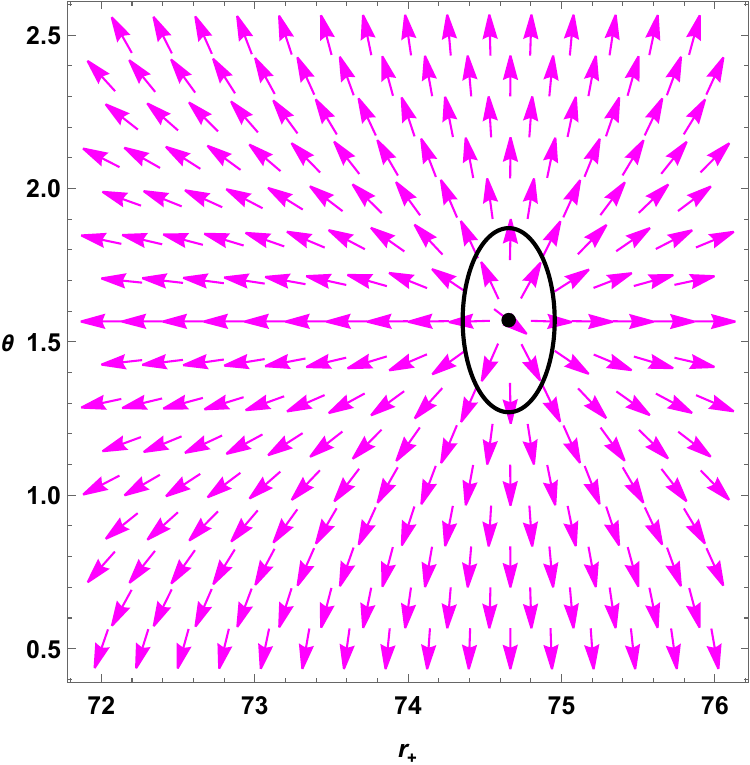}
		\caption{}
		\label{c8b}
	\end{subfigure}\hspace{0.2cm} 
\begin{subfigure}{0.35\textwidth}
		\includegraphics[width=\linewidth]{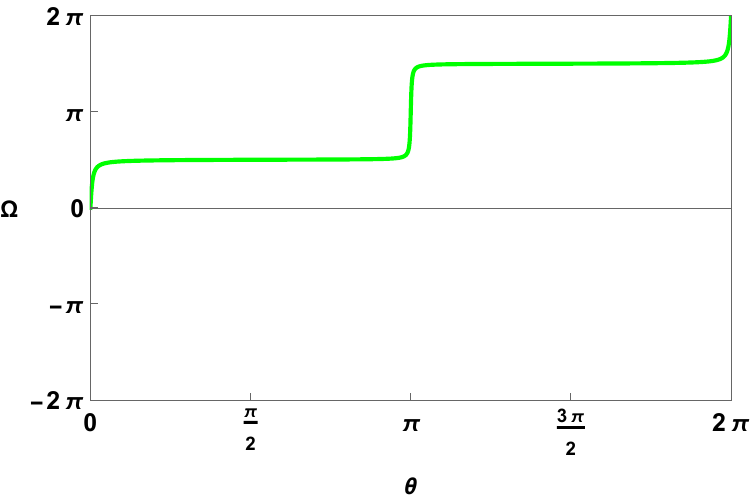}
		\caption{}
		\label{c8c}
	\end{subfigure}
\caption{Fig. \protect\ref{c6a} is the $\protect\tau$ vs $r_+$ plot for $%
\protect\alpha=0.15$, $q=2$, $b=1$, $\Lambda = -0.01$, and $k=0$. Fig. 
\protect\ref{c6b} is the vector plot in the $r-\protect\theta$ plane for $%
\protect\tau=30$. As it is clear from the vector plot, the zero point is
found to be at $(74.6558, \frac{\protect\pi}{2})$. In Fig. \protect\ref{c6c}%
, the green contour represents the winding number for the black hole, which
equals $1$.}
\label{c8}
\end{figure}

\begin{figure}[h!]
\centering
\begin{subfigure}{0.28\textwidth}
		\includegraphics[width=\linewidth]{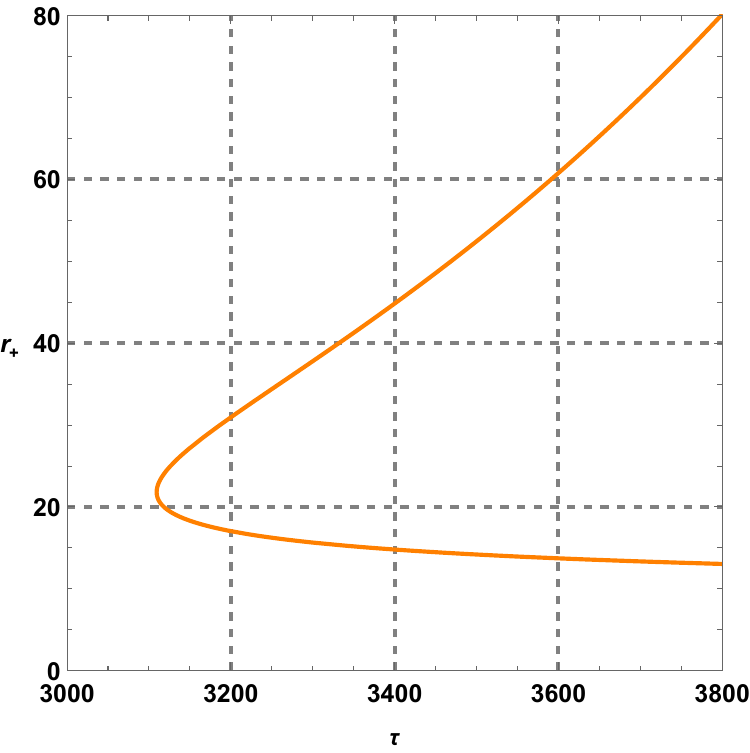}
		\caption{}
		\label{c9a}
	\end{subfigure}
\hspace{0.5cm} 
\begin{subfigure}{0.28\textwidth}
		\includegraphics[width=\linewidth]{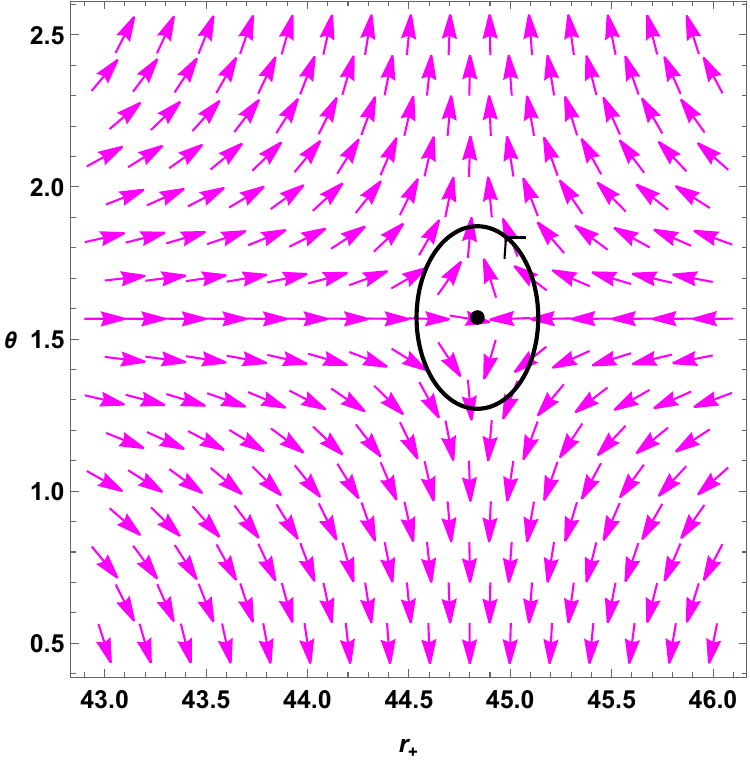}
		\caption{}
		\label{c9b}
	\end{subfigure}\newline
\begin{subfigure}{0.28\textwidth}
		\includegraphics[width=\linewidth]{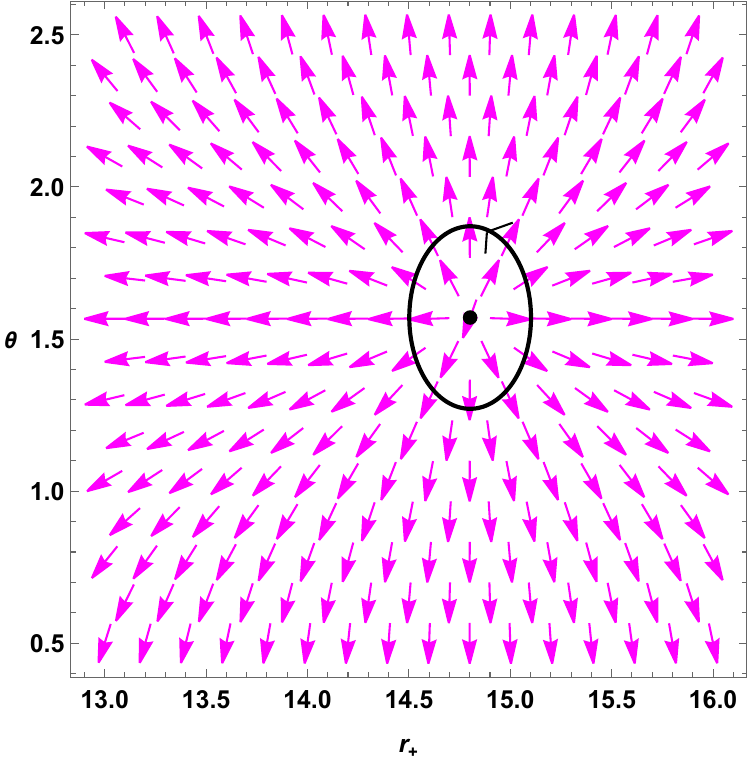}
		\caption{}
		\label{c9c}
	\end{subfigure}
\hspace{0.5cm} 
\begin{subfigure}{0.35\textwidth}
		\includegraphics[width=\linewidth]{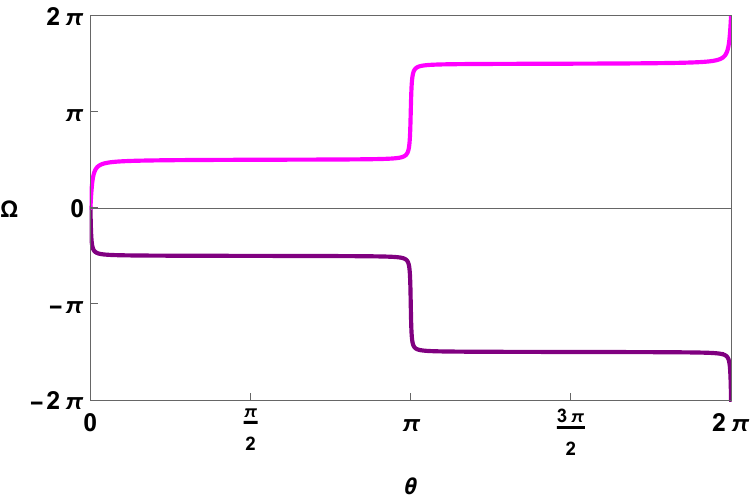}
		\caption{}
		\label{c9d}
	\end{subfigure}
\caption{Fig. \protect\ref{c7a} is the $\protect\tau$ vs $r_+$ plot for $%
\protect\alpha=0.5$, $q=2$, $b=1$, $\Lambda=-0.01$, and $k=0$. Fig. \protect
\ref{c7b} and Fig. \protect\ref{c7c} are the vector plot in the $r-\protect%
\theta$ plane for $\protect\tau=3400$. As it is clear from the vector plots,
the zero points are found to be at $(14.8022, \frac{\protect\pi}{2})$ in the
small black hole branch and $(44.838, \frac{\protect\pi}{2})$ in the large
black hole branch. In Fig. \protect\ref{c7d}, the purple contour represents
the winding number for a large black hole branch and the magenta contour
represents the same for a small black hole branch.}
\label{c9}
\end{figure}

Here also two topological classes of black holes with topological charge $0$
and $+1$ are found depending on the value of $\alpha$. It is important to
note that for $k=0$ case also, only negative values of $\Lambda$ are allowed
due to positive temperature conditions. Figs. \ref{c8} and \ref{c9}
demonstrate that the different topological classes exhibited by these black
holes.


\subsection{\textbf{Fixed potential ($\protect\phi $) ensemble}}

In the fixed potential $\phi$ ensemble, the potential $\phi$ is held
constant, serving as the conjugate to the charge $q$. The expression for $%
\phi$ is derived in the following form 
\begin{equation}
\phi =\frac{\partial M}{\partial q}=\frac{q\left( \frac{b}{r_{+}}\right) {}^{%
\frac{4\alpha ^{2}}{\mathcal{K}_{1,1}}}}{4\pi r_{+}},  \label{phi}
\end{equation}%
where $q$ can be extracted from the Eq. (\ref{phi}) as 
\begin{equation}
q=4\pi \phi \left( \frac{b}{r_{+}}\right) {}^{-\frac{4\alpha ^{2}}{\mathcal{K%
}_{1,1}}}r_{+}.  \label{charge}
\end{equation}

Substituting $q$ from Eq. (\ref{charge}) in the expression of mass in Eq. (%
\ref{Mass}), new mass in this ensemble is obtained as 
\begin{equation}
\tilde{M}=\frac{\left( \frac{b}{r_{+}}\right) {}^{\frac{4\alpha ^{2}}{%
\mathcal{K}_{1,1}}}}{8\pi }\left( 16\pi ^{2}\phi ^{2}r_{+}+\frac{\mathcal{K}%
_{1,1}\Lambda r_{+}^{3}\left( \frac{b}{r_{+}}\right) {}^{\frac{4\alpha ^{2}}{%
\mathcal{K}_{1,1}}}}{\mathcal{K}_{1,-3}}-\frac{kr_{+}}{\mathcal{K}_{1,-1}}%
\right) .  \label{MMG}
\end{equation}

The free energy formula in this ensemble is modified as follows 
\begin{equation}
\mathcal{F}=\tilde{M}-\frac{S}{\tau }-q\phi ,  \label{FF}
\end{equation}%
using Eqs. (\ref{entropy}), (\ref{charge}), (\ref{MMG}), and (\ref{FF}) the
free energy is obtained as 
\begin{equation}
\mathcal{F}=-\frac{\left( \frac{b}{r_{+}}\right) {}^{\frac{2\alpha ^{2}}{%
\mathcal{K}_{1,1}}}r_{+}^{2}}{4\tau }+\frac{\left( -\frac{k\left( \frac{b}{%
r_{+}}\right) {}^{\frac{4\alpha ^{2}}{\mathcal{K}_{1,1}}}}{\mathcal{K}_{1,-1}%
}+\frac{\mathcal{K}_{1,1}\mathcal{K}_{1,-1}\Lambda r_{+}^{2}}{\mathcal{K}%
_{1,-1}\mathcal{K}_{1,-3}}\left( \frac{b}{r_{+}}\right) {}^{\frac{8\alpha
^{2}}{\mathcal{K}_{1,1}}}-\frac{16\pi ^{2}\phi ^{2}}{\left( \frac{b}{r_{+}}%
\right) {}^{\frac{4\alpha ^{2}}{\mathcal{K}_{1,1}}}}\right) r_{+}}{8\pi }.
\end{equation}

Following the same prescription shown in the previous sections, the zero
points of the $\phi ^{r}$ component are obtained as 
\begin{equation}
\tau =\frac{4\pi r_{+}}{\frac{\mathcal{K}_{3,-1}k\left( \frac{b}{r_{+}}%
\right) {}^{\frac{2\alpha ^{2}}{\mathcal{K}_{1,1}}}}{\mathcal{K}_{1,-1}}-%
\frac{\mathcal{K}_{1,1}\mathcal{K}_{5,-3}\Lambda r_{+}^{2}\left( \frac{b}{%
r_{+}}\right) {}^{\frac{6\alpha ^{2}}{\mathcal{K}_{1,1}}}}{\mathcal{K}_{1,-3}%
}-\frac{16\pi ^{2}\mathcal{K}_{5,1}\phi ^{2}}{\left( \frac{b}{r_{+}}\right)
{}^{\frac{6\alpha ^{2}}{\mathcal{K}_{1,1}}}}}.  \label{gtau}
\end{equation}

\subsubsection{\textbf{For elliptic ($k=+1$) curvature hypersurface}}

\begin{figure}[h]
\centering
\begin{subfigure}{0.28\textwidth}
		\includegraphics[height=5cm,width=5cm]{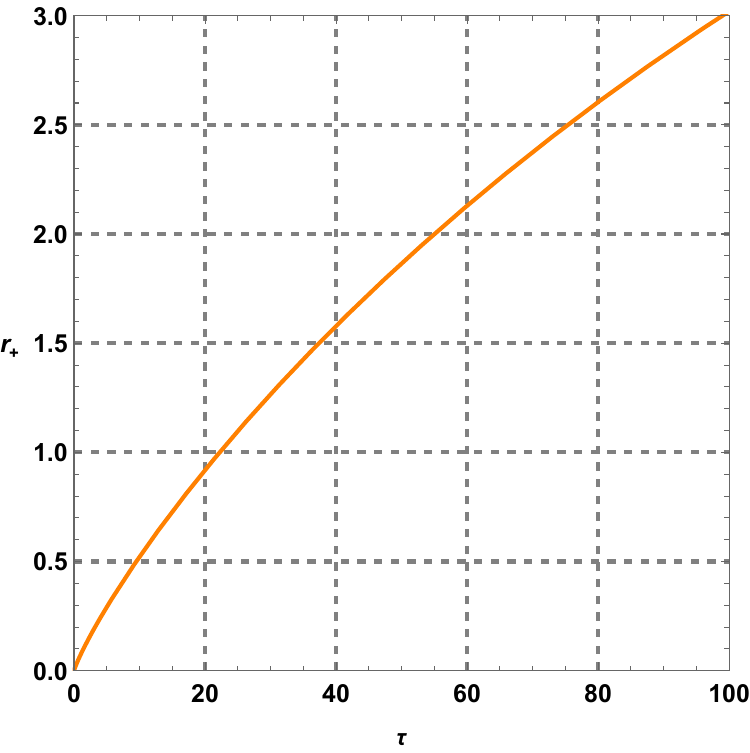}
		\caption{}
		\label{c10a}
	\end{subfigure}
\begin{subfigure}{0.28\textwidth}
		\includegraphics[height=5cm,width=5cm]{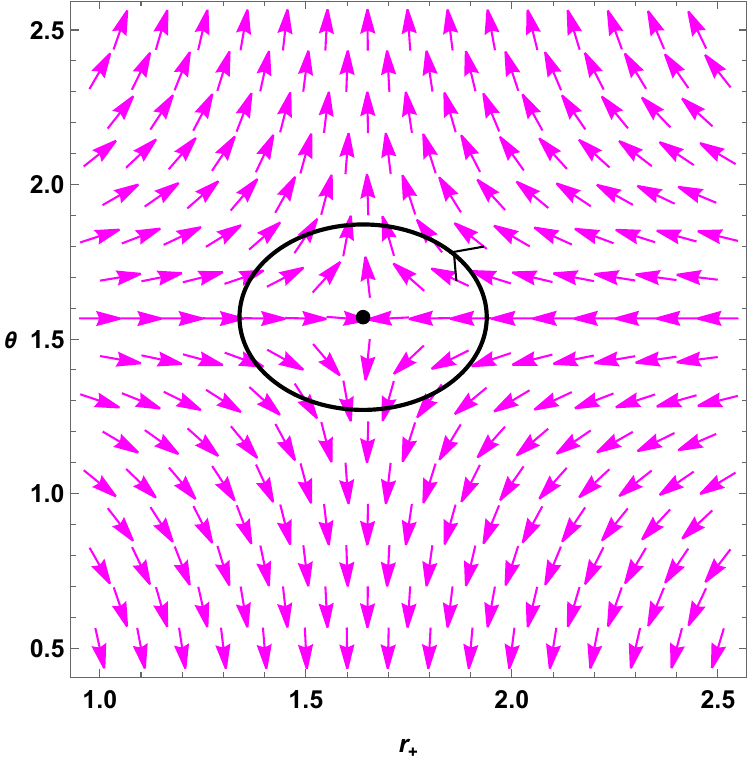}
		\caption{}
		\label{c10b}
	\end{subfigure} 
\begin{subfigure}{0.28\textwidth}
		\includegraphics[height=5cm,width=5cm]{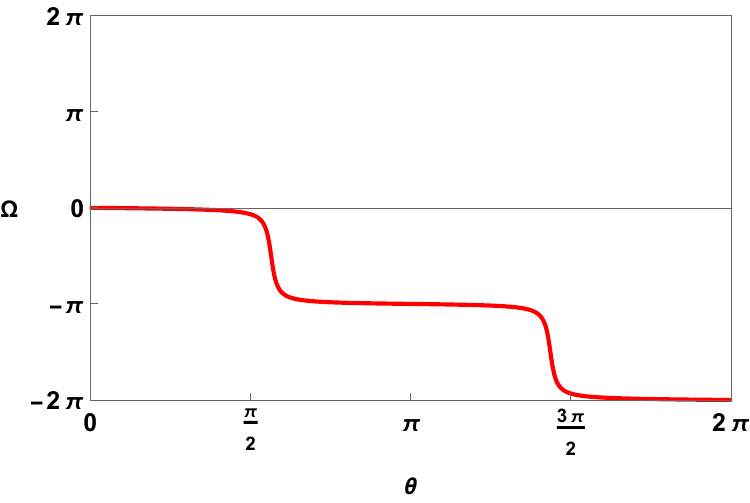}
		\caption{}
		\label{c10c}
	\end{subfigure}
\caption{Topological charge of charged dilatonic black holes ($k=+1$) in
fixed potential ensemble when positive values of $\Lambda $ are considered.}
\label{gc1}
\end{figure}

For elliptic curvature hypersurface case, we substitute $k=+1$ in the
expression Eq. (\ref{gtau}). For $\Lambda>0$, we found a new topological
class of topological charge $W=-1$ in this ensemble. The graph in Fig. \ref%
{c10a} depicts the relationship between $\tau$ and $r_{+}$. The parameters
used for this plot are: $\phi=0.02$, $\alpha =0.25$, $b=0.2$, $\Lambda =0.01$%
, and $k=1$. Only a single branch of the black hole is visible. In Fig. \ref%
{c10b}, a vector plot is presented for the components $\phi ^{r}$ and $\phi
^{\theta }$ with $\tau=40$. The zero points of the vector field are at $%
r_{+}=1.6390$. Fig. \ref{c10c} shows that the topological charge for $%
r_{+}=1.6390$ is $-1$, as indicated by the red line.

\begin{figure}[h]
\centering
\begin{subfigure}{0.28\textwidth}
		\includegraphics[width=\linewidth]{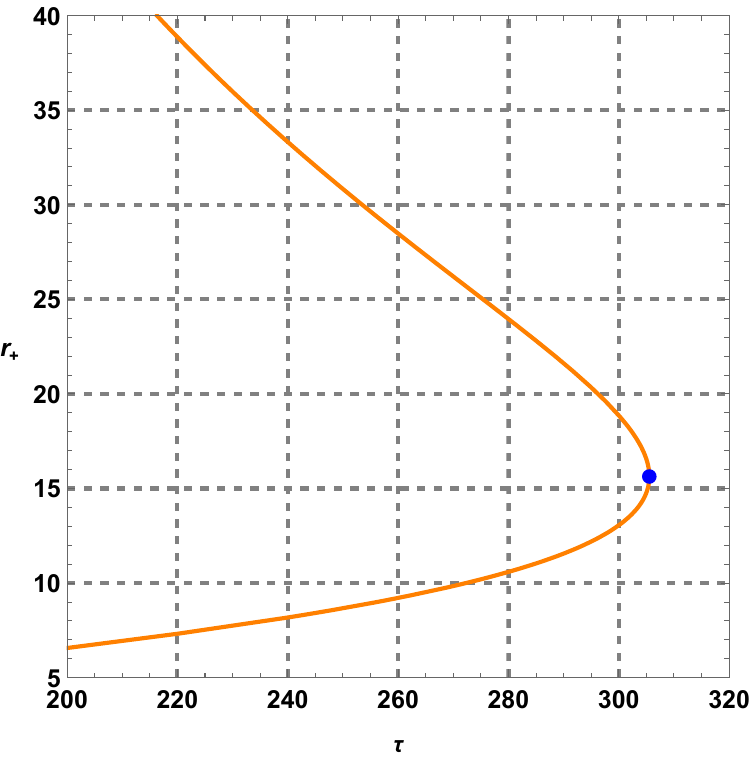}
		\caption{}
		\label{c11a}
	\end{subfigure}
\begin{subfigure}{0.28\textwidth}
		\includegraphics[width=\linewidth]{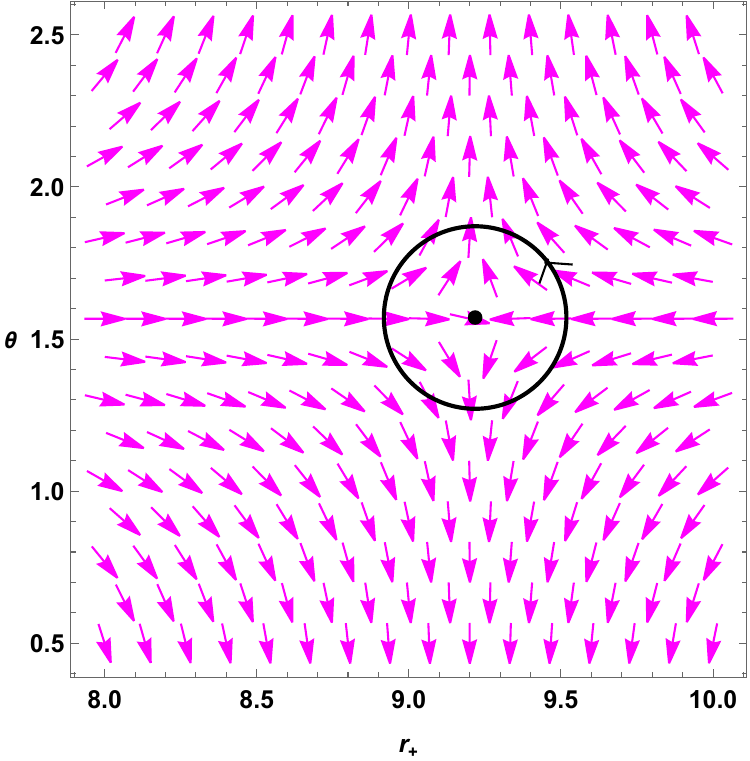}
		\caption{}
		\label{c11b}
	\end{subfigure} \hspace{0.6cm} 
\begin{subfigure}{0.28\textwidth}
		\includegraphics[width=\linewidth]{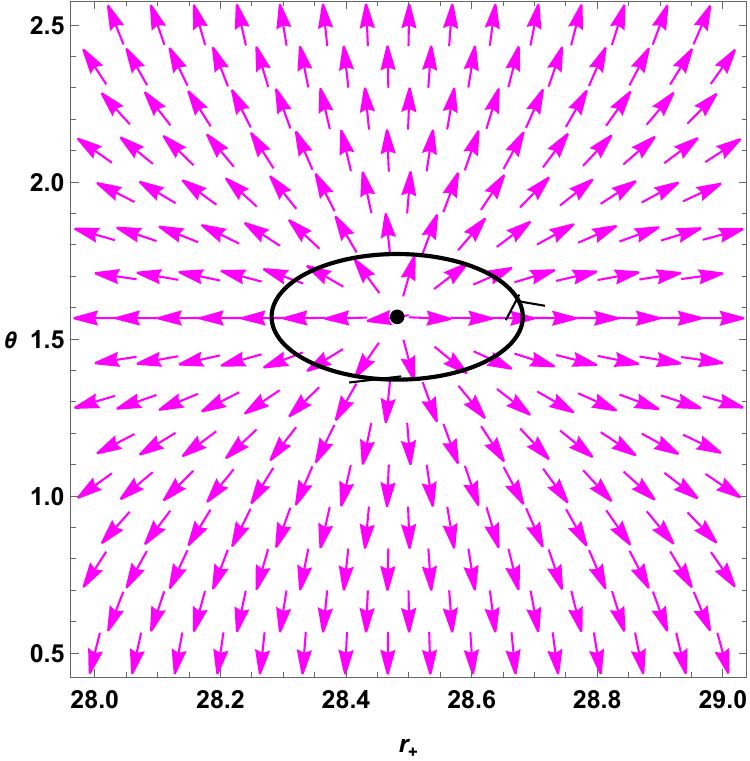}
		\caption{}
		\label{c11c}
	\end{subfigure}
\begin{subfigure}{0.3\textwidth}
		\includegraphics[width=\linewidth]{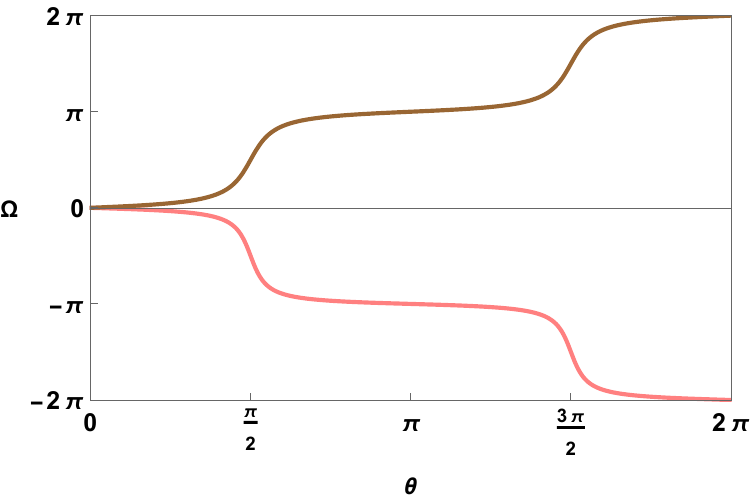}
		\caption{}
		\label{c11d}
	\end{subfigure}
\hspace{0.6cm} 
\begin{subfigure}{0.3\textwidth}
		\includegraphics[width=\linewidth]{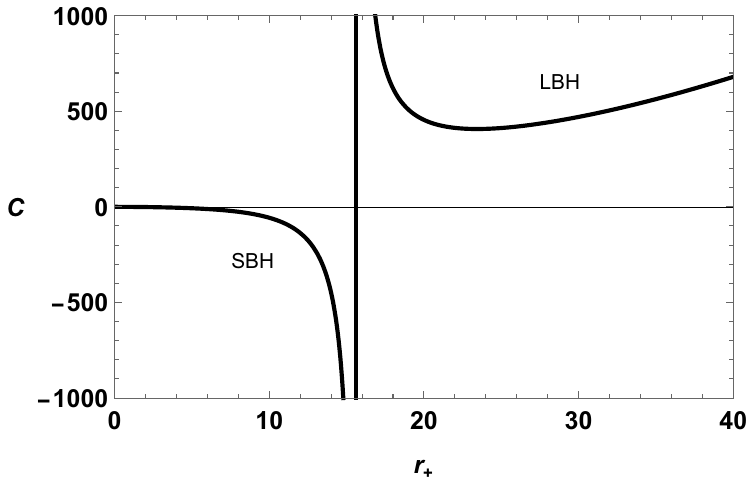}
		\caption{}
		\label{c11e}
	\end{subfigure}
\caption{Fig. \protect\ref{c11a} is the $\protect\tau $ vs $r_{+}$ of
topological charged dilatonic black holes with elliptical curvature
hypersurface $(k=+1)$ in the fixed potential ensemble when the negative
value of $\Lambda $ is considered. We have used $\protect\phi =0.02$, $%
\protect\alpha=0.25$, $b=0.2$, $k=+1$, and $\Lambda =-0.01$ for this plot.
Fig. \protect\ref{c11b} and Fig. \protect\ref{c11c} are the vector plot in
the $r_{+}-\protect\theta $ plane for $\protect\tau =260$. As it is clear
from the vector plots, the zero points are found to be at $(9.2178,\frac{%
\protect\pi }{2})$ in the small black hole branch and $(28.4815,\frac{%
\protect\pi }{2})$ in the large black hole branch. In Fig. \protect\ref{c7d}%
, the brown contour represents the winding number for the large black hole
branch, and the pink contour represents the same for the small black hole
branch. As Fig. \protect\ref{c11e} depicts, the small black holes are
unstable, whereas the large black holes are stable. In addition, the Davies
point is located at $r_{+}=15.6359$.}
\label{c11}
\end{figure}

Next, we consider the negative values of $\Lambda $. In this scenario,
although the topological charge is still zero, the stability pattern of
black hole branches changes. In previous cases with a topological charge of
zero, the small black hole branch was stable with a winding number of $+1$,
while the large black hole branch was unstable with a winding number of $-1$%
. However, as depicted in Fig. \ref{c11a}, despite the presence of two
branches in the $\tau$ vs $r_{+}$ plot, the smaller black hole now has a
winding number of $-1$, and the large black hole branch has a winding number
of $+1$, represented by the pink and brown solid lines in Fig. \ref{c11d}.
These plots were generated using the parameters $\phi =0.02$, $\alpha =0.25$%
, $b=0.2$, $k=1$, $\Lambda=-0.01$, and $\tau =260$. Unlike the fixed charge
ensemble, here we get a generation point which is located at $r_{+}=15.6359$
for the plot Fig. \ref{c11a} represented by the blue dot. In Fig. \ref{c11e}%
, we explicitly demonstrate the change in the stability pattern of the black
hole branches by plotting the specific heat $C$ against $r_{+}$. To evaluate 
$C$ we have used the following formula 
\begin{equation}
C=\frac{dM}{dT},  \label{CC}
\end{equation}%
where $T=\frac{1}{\tau }$ for thermodynamically stable black hole branch $C$
is positive, and it is negative for thermodynamically unstable black hole
branch. As Fig. \ref{c11e} depicts, the small black hole branch has a
negative specific heat and the large black hole branch has a positive
specific heat, which explains their stability. The Davies point is located
at $r_{+}=15.6359$.

\begin{figure}[h]
\centering
\begin{subfigure}{0.28\textwidth}
		\includegraphics[width=\linewidth]{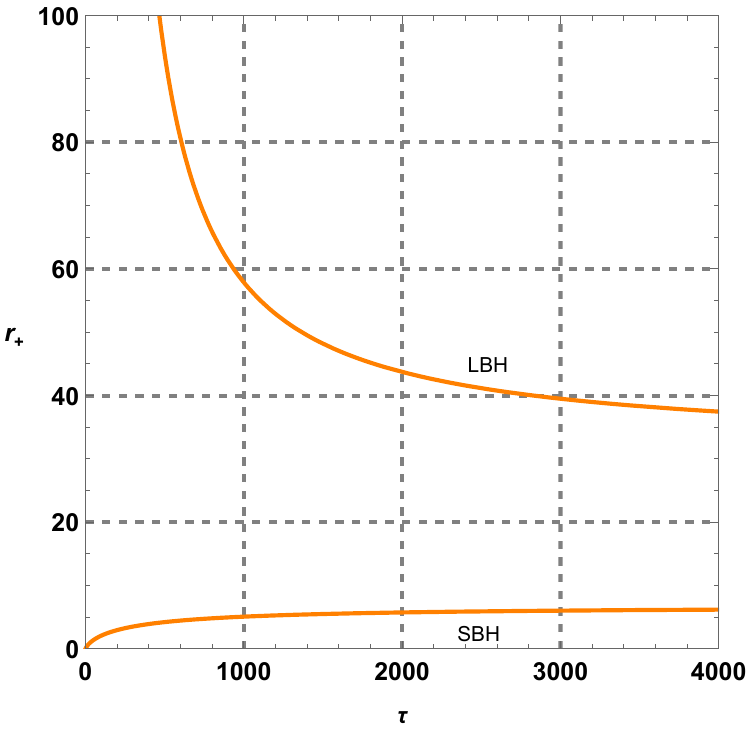}
		\caption{}
		\label{c12a}
	\end{subfigure}
\begin{subfigure}{0.28\textwidth}
		\includegraphics[width=\linewidth]{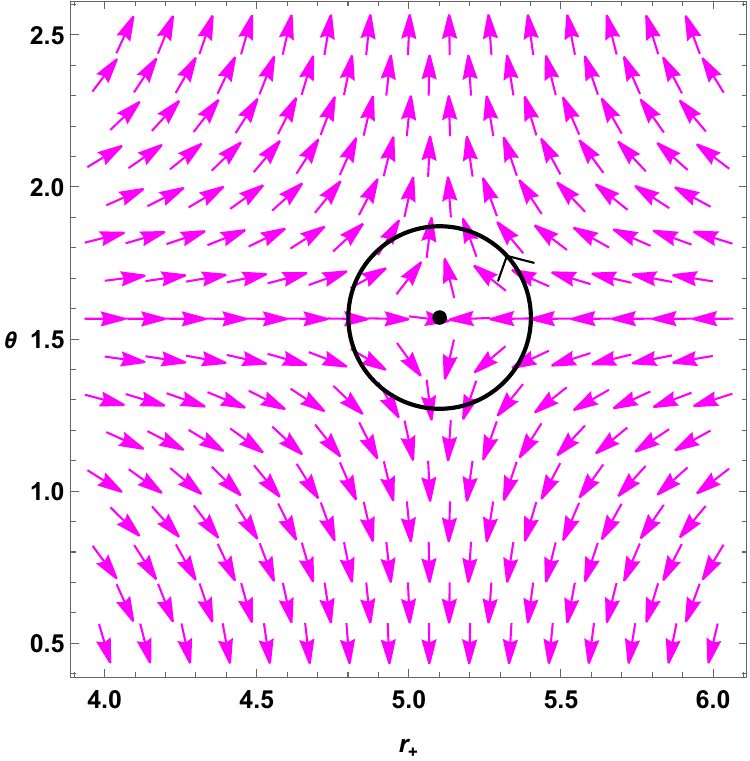}
		\caption{}
		\label{c12b}
	\end{subfigure}\newline
\begin{subfigure}{0.28\textwidth}
	\includegraphics[width=\linewidth]{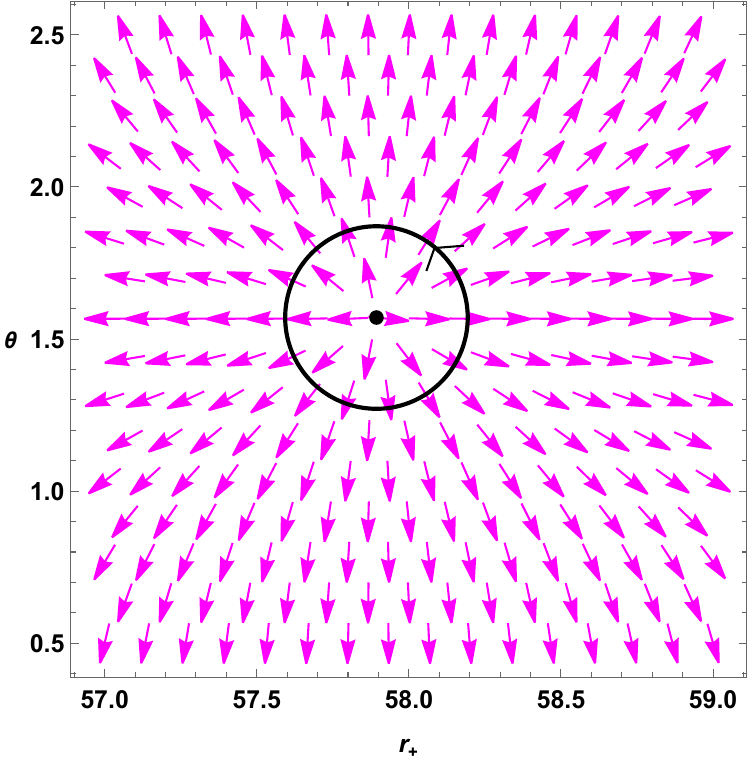}
	\caption{}
	\label{c12c}
\end{subfigure} 
\begin{subfigure}{0.3\textwidth}
		\includegraphics[width=\linewidth]{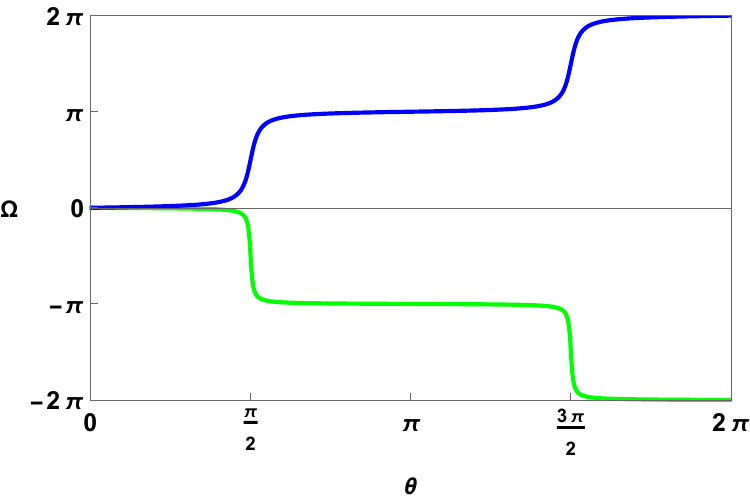}
		\caption{}
		\label{c12d}
	\end{subfigure} 
\caption{Fig. \protect\ref{c12a} is the $\protect\tau$ vs $r_+$ of charged
dilatonic black holes with elliptical curvature hypersurface $(k=+1)$ in
fixed potential ensemble for $\protect\phi=0.02$, $\protect\alpha=0.25$, $%
b=0.2$, $k=+1$, and $\Lambda=-0.01$. Two black hole branches are observed: a
large black hole branch (LBH) and a small black hole branch (SBH). Fig. 
\protect\ref{c12b} and Fig. \protect\ref{c12c} are the vector plot in the $%
r_{+}-\protect\theta$ plane for $\protect\tau=1000$. As it is clear from the
vector plots, the zero points are found to be at $(5.1010, \frac{\protect\pi%
}{2})$ in the small black hole branch and $(57.8935, \frac{\protect\pi}{2})$
in the large black hole branch. In Fig. \protect\ref{c12d}, the blue contour
represents the winding number for the large black hole branch and the green
contour represents the same for the small black hole branch.}
\label{12}
\end{figure}

Another interesting scenario is observed when we increase the value of $%
\alpha$ while keeping the other parameters constant at $\phi=0.02$, $b=0.2$, 
$k=+1$, and $\Lambda =-0.01$. For example, when we set $\alpha =0.3$, we
observe two discontinuous black hole branches, as shown in Fig. \ref{c12a}.
Besides these two black hole branches, the other branches fall into the
negative temperature zone and have therefore been omitted. For the small
black hole branch, the winding number is $-1$, while for the large black
hole branch, the winding number is $+1$. Consequently, the topological
charge is $W=-1+1=0$.

For this analysis, our concluding remark is as follows: the charged
dilatonic black hole with an elliptic ($k=+1$) curvature hypersurface has
two topological classes, $W=-1$ and $W=0$, in the fixed potential ensemble,
depending on the sign of $\Lambda$. Previously, we also identified a
topological class with a topological charge of $0$ in the fixed charge
ensemble. However, this class is distinct from the $W=0$ class found in the
fixed potential ensemble, as the local topology of these classes differs.
Although both classes share the same global topology with a topological
charge of $W=0$, the stability pattern and the local winding number of the
black hole branches differ. Interestingly, in this ensemble, apart from the
sign of $\Lambda$, none of the thermodynamic parameters significantly
influence the topological charge.


\subsubsection{\textbf{For hyperbolic ($k=-1$) curvature hypersurface}}

\begin{figure}[h!]
\centering
\begin{subfigure}{0.28\textwidth}
		\includegraphics[width=\linewidth]{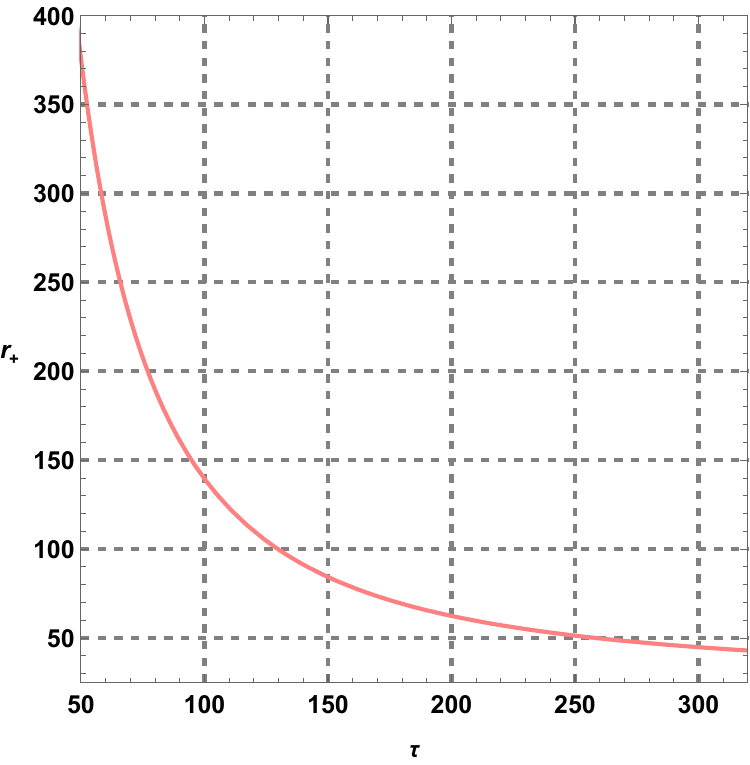}
		\caption{}
		\label{c13a}
	\end{subfigure}
\begin{subfigure}{0.28\textwidth}
		\includegraphics[width=\linewidth]{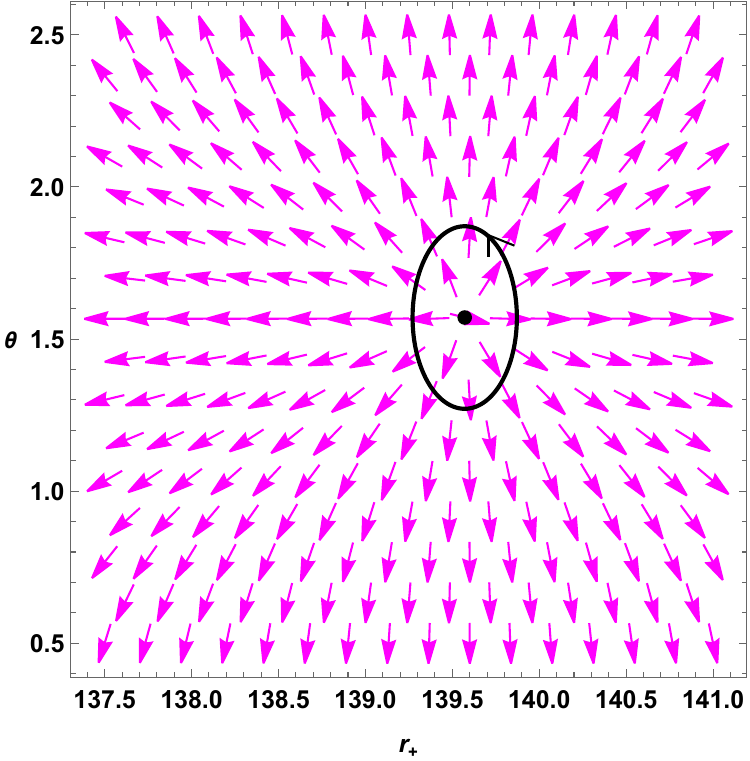}
		\caption{}
		\label{c13b}
	\end{subfigure} 
\begin{subfigure}{0.32\textwidth}
		\includegraphics[width=\linewidth]{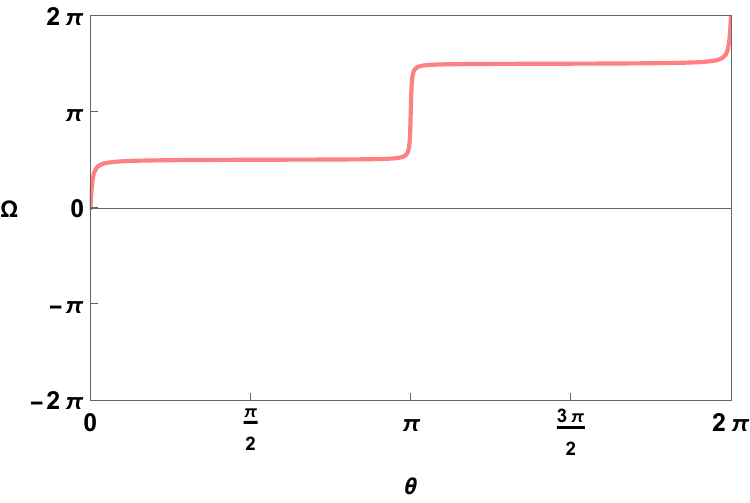}
		\caption{}
		\label{c13c}
	\end{subfigure}
\caption{Fig. \protect\ref{c13a} is the $\protect\tau$ vs $r_+$ of the
charged dilatonic black holes with hyperbolic curvature hypersurface $(k=-1)$
in fixed potential ensemble for $\protect\phi=0.02$, $\protect\alpha=0.25$, $%
b=0.2$, $k=-1$, and $\Lambda=-0.01$, where a single black hole branch is
observed. Fig. \protect\ref{c13b} is the vector plot in the $r_{+}-\protect%
\theta$ plane for $\protect\tau=100$. As it is clear from the vector plot,
the zero point is found to be at $(139.5713, \frac{\protect\pi}{2})$. In
Fig. \protect\ref{c13c}, the pink contour represents the winding number for
the black hole, which is found to be $W=+1$.}
\label{13}
\end{figure}

To evaluate the effect of hyperbolic curvature on the under-studying system,
we substitute $k=-1$ in Eq. (\ref{gtau}). We find only a single black hole
branch (see Fig. \ref{c13a}). The topological charge is $+1$. In this
ensemble, we do not get the topological class with topological charge $W=0$
as we have found in the fixed charge ensemble for the black hole with
hyperbolic ($k=-1$) curvature hypersurface. Here also, the topological
charge is invariant with thermodynamic parameters.

\subsubsection{\textbf{For flat ($k=0$) curvature hypersurface}}

\begin{figure}[h!]
\centering
\begin{subfigure}{0.28\textwidth}
		\includegraphics[width=\linewidth]{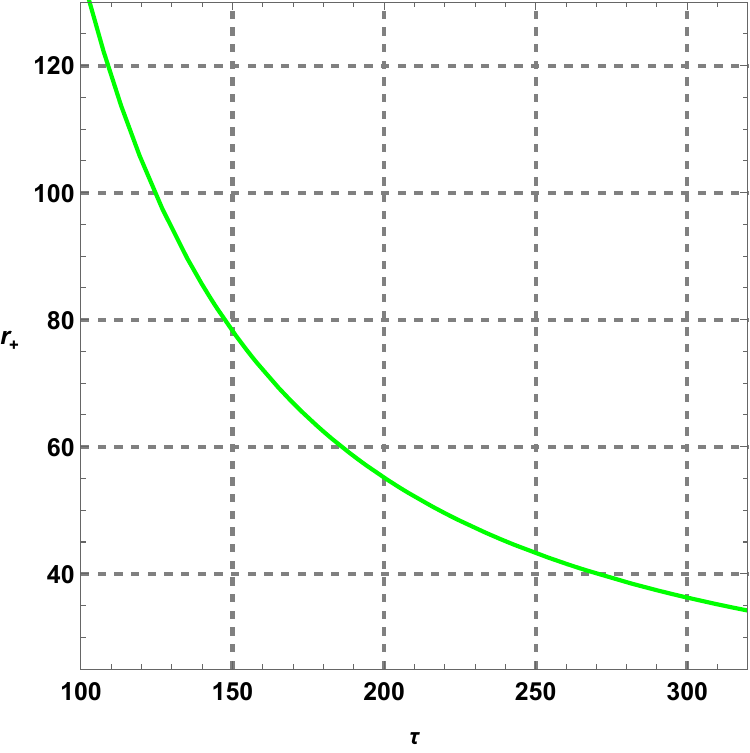}
		\caption{}
		\label{c14a}
	\end{subfigure}
\begin{subfigure}{0.28\textwidth}
		\includegraphics[width=\linewidth]{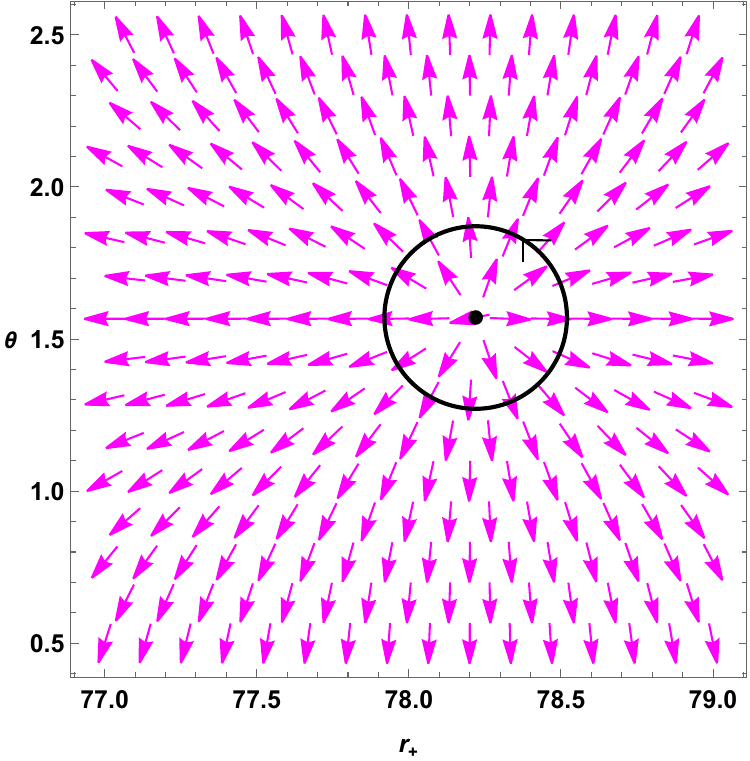}
		\caption{}
		\label{c14b}
	\end{subfigure} 
\begin{subfigure}{0.32\textwidth}
		\includegraphics[width=\linewidth]{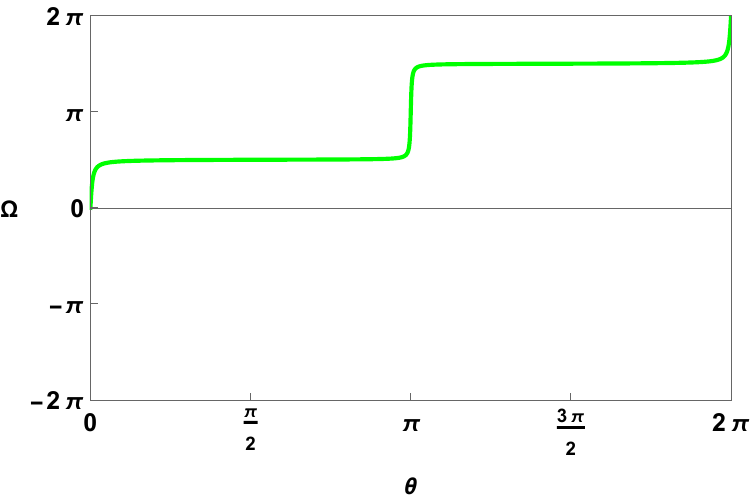}
		\caption{}
		\label{c14c}
	\end{subfigure}
\caption{Fig. \protect\ref{c14a} is the $\protect\tau$ vs $r_+$ of charged
dilatonic black holes with flat curvature hypersurface ($k=0$) in fixed
potential ensemble for $\protect\phi= 0.02$, $\protect\alpha=0.25$, $b=0.2$, 
$k=0$, $\Lambda=-0.01$ where a single black hole branch is observed. Fig. 
\protect\ref{c13b} is the vector plot in the $r_{+}-\protect\theta $ plane
for $\protect\tau=150$. As it is clear from the vector plot, the zero point
is found to be at $(78.2196, \frac{\protect\pi}{2})$. In Fig. \protect\ref%
{c13c}, the green contour represents the winding number for the black hole,
which is found to be $W=+1$.}
\label{14}
\end{figure}

We substitute $k=0$ in Eq. (\ref{gtau}). In this case, we find only a single
black hole branch, as shown in Fig. \ref{c14a}. The topological charge is
again determined to be $+1$. For these types of black holes also, we do not
encounter the topological class with a topological charge of $W=0$, as we
found in the fixed charge ensemble. Additionally, the topological charge
remains invariant with respect to the thermodynamic parameters.

Our findings are reported in Table. \ref{tableI}, based on the effect of the
topological constant $k$. In other words, the Table. \ref{tableI}
categorizes the results for elliptical ($k=1$), flat ($k=0$), and hyperbolic
($k=-1$) curvature hypersurfaces.

\begin{table*}[htb!]
\caption{}
\label{tableI}\centering
\begin{tabular}{cccccc}
\hline\hline
ensembles & $k$ & $\Lambda$ & topological charge \hspace{0.3cm} & generation
point \hspace{0.3cm} & annihilation point \\ \hline\hline
& $k=+1$ & $\Lambda>0$ & $0$ & $1$ & $0$ \\ 
fixed $q$ ensemble &  & $\Lambda<0$ & $1$ or $0$ & $1$ or $0$ & $1$ or $0$
\\ 
& $k=-1$ & $\Lambda<0$ & $0$ or $1$ & $1$ or $0$ & $0$ \\ 
& $k=0$ & $\Lambda<0$ & $0$ or $1$ & $0$ or $1$ & $0$ \\ \hline
& $k=+1$ & $\Lambda>0$ & $-1$ & $0$ & $0$ \\ 
fixed $\phi $ ensemble &  & $\Lambda<0$ & $0$ & $0$ & $1$ \\ 
& $k=-1$ & $\Lambda<0$ & $1$ & $0$ & $0$ \\ 
& $k=0$ & $\Lambda<0$ & $1$ & $0$ & $0$ \\ \hline\hline
\end{tabular}%
\end{table*}

\section{Conclusion}

In this paper, we first explored the concept of topological charged
dilatonic black holes, which were black holes in dilaton gravity with the
presence of the Maxwell field. We calculated the Kretschmann scalar and
found that in the presence of the dilaton field, the asymptotic behavior of
the spacetime changed. Indeed, black hole spacetimes were neither
asymptotically flat nor (A)dS. We then examined the impact of the
topological constant ($k$) on the event horizon. Our findings indicated that
black holes with a large size corresponded to a negative value of the
topological constant, i.e., $k=-1$. Our analysis is presented in Figure. \ref%
{Fig2}, which revealed that for $k=+1$ and $k=0$, the large black holes had
a small value of $\alpha$ and a large value of $b$. However, in the case of $%
k=-1$, the large black holes exhibited large values of $\alpha$ and small
values of $b$, which was different from the previous two cases.

In section III, we calculated the thermodynamic and conserved quantities for
topological charged dilatonic black holes to study their thermodynamic
properties. Additionally, the extracted thermodynamic quantities satisfied
the first law of thermodynamics.

In Section IV, we explored the thermodynamic topology of these black holes
using the off-shell free energy method. We studied two types of
thermodynamic ensembles: the fixed $q$ ensemble and the fixed $\phi$
ensemble. In the fixed $q$ ensemble, we first considered the case of
elliptic curvature ($k=+1$) hypersurfaces, taking both $\Lambda>0$ and $%
\Lambda<0$ into consideration. For a positive value of $\Lambda$, we
identified a single topological class $W=0$ with one annihilation point. The
topological charge remained invariant despite variations in all
thermodynamic parameters of dilaton gravity. Upon shifting $\Lambda$ to
negative values, we discovered two topological classes $W=+1$ and $W=0$.
Notably, the topological class of the black hole was dependent on the
parameter $\alpha$ of dilaton gravity. Additionally, we found one
annihilation point and either one or zero generation points in this
scenario. For the cases of flat ($k=0$) and hyperbolic ($k=-1$) curvature
hypersurfaces, we identified two topological classes, $W=+1$ and $W=0$, with
either one or zero annihilation points. Here, the topological charge also
showed dependence on the parameter $\alpha$ of dilaton gravity for
topological constants $k=0$ and $k=-1$.

In the fixed potential $\phi$ ensemble, we observed a topological class $%
W=-1 $ for $\Lambda>0$, and a topological class $W=0$ with a generation
point for $\Lambda<0$. Notably, the local thermodynamic topology changes
significantly in the fixed $\phi$ ensemble. Previously, in the fixed charge (%
$q$) ensemble, we also identified a $W=0$ topological class, but its local
thermodynamic topology contrasts with that found in the fixed $\phi$
ensemble's $W=0$ class. Despite having the same topological charge, their
winding numbers revealed opposite local topologies.

Similarly, in the fixed charge ensemble for flat ($k=0$) and hyperbolic ($%
k=-1$) curvature hypersurfaces, we found only one topological class, $W=+1$,
which lacked annihilation and generation points. Conversely, in the fixed $%
\phi$ ensemble, the topological charge remained invariant despite variations
in all thermodynamic parameters of dilaton gravity.

In summary, we identified three distinct topological classes, $W=+1$, $0$, $%
-1$, for charged topological black holes with $\Lambda>0$ and $\Lambda<0$ in
dilaton gravity across different ensembles, contingent upon the values of
the topological constant $k$, parameter $\alpha$, and the sign of the
cosmological constant $\Lambda$. These findings are synthesized in Table. %
\ref{tableII}, which also includes thermodynamic topological properties of
Reissner-Nordstrom (RN) black holes for comparison (Ref. \cite{TheTo2}).

\begin{table*}[htb!]
\caption{}
\label{tableII}\centering
\begin{tabular}{cccc}
\hline\hline
Ensembles & Topological Properties\hspace{0.3cm} & topological charged
dilatonic black holes \hspace{0.3cm} & RN black holes \\ \hline\hline
& Topological Charge & $0$ or $1$ & $0$ or $1$ \\ 
Fixed charge $(q)$ ensemble & Generation Point & $0$ or $1$ & $0$ or $1$ \\ 
& Annihilation Point & $0$ or $1$ & $0$ or $1$ \\ \hline
& Topological Charge & $-1$, $0$ or $1$ & $-1$, $0$ \\ 
Fixed potential $(\phi )$ ensemble & Generation Point & $1$ or $0$ & $1$ or $%
0$ \\ 
& Annihilation Point & $1$ or $0$ & $0$ \\ \hline\hline
\end{tabular}%
\end{table*}

\begin{acknowledgements}
BEP would like to thank the University of Mazandaran. BH would like to thank
DST-INSPIRE, Ministry of Science and Technology fellowship program, Govt. of
India for awarding the DST/INSPIRE Fellowship [IF220255] for financial
support.
\end{acknowledgements}

\end{document}